\documentclass[journal]{IEEEtran}
\IEEEoverridecommandlockouts
\usepackage{cite}
\usepackage{amsmath,amssymb,amsfonts}
\usepackage{algorithmic}
\usepackage{graphicx}
\usepackage{textcomp}
\usepackage{fancyhdr}
\usepackage{xcolor}
\usepackage{lipsum}
\usepackage{enumitem} 
\usepackage{algorithm}
\usepackage{algorithmic}

\usepackage{array}        % needed for newcolumntype
\usepackage{multirow}     % for \multirow
\usepackage{booktabs}     % nicer rules (optional)
\usepackage{makecell}     % line breaks in headers (optional)

\newcolumntype{P}[1]{>{\raggedright\arraybackslash}p{#1}}

\def\BibTeX{{\rm B\kern-.05em{\sc i\kern-.025em b}\kern-.08em
    T\kern-.1667em\lower.7ex\hbox{E}\kern-.125emX}}

\begin{document}
%\history{Date of publication xxxx 00, 0000, date of current version xxxx 00, 0000.}
%\doi{xx.xxxx/ACCESS.xxxx.xxxxxxx}

\title{Approximate Bilevel Graph Structure Learning for Histopathology Image Classification}

\author{Sudipta~Paul,~\IEEEmembership{Member,~IEEE},
Amanda~W.~Lund, George~Jour, Iman~Osman, and B\"ulent~Yener,~\IEEEmembership{Fellow,~IEEE}%
\thanks{Corresponding author: Sudipta Paul (pauls5@rpi.edu).}%
\thanks{S. Paul is with the \emph{Department of Electrical, Computer, and Systems Engineering}, Rensselaer Polytechnic Institute (RPI), Troy, NY 12180, USA.}%
\thanks{A. W. Lund and G. Jour are with the \emph{Department of Pathology}, and with the \emph{Ronald O. Perelman Department of Dermatology}, NYU Grossman School of Medicine, New York, NY 10016, USA.}%
\thanks{I. Osman is with the \emph{Ronald O. Perelman Department of Dermatology}, the \emph{Department of Medicine}, and the \emph{Department of Urology}, NYU Grossman School of Medicine, New York, NY 10016, USA.}%
\thanks{B. Yener is with the \emph{Department of Electrical, Computer, and Systems Engineering}, and with the \emph{Department of Computer Science}, RPI, Troy, NY 12180, USA.}%
}

% <<< This makes the abstract + keywords span both columns >>>
\IEEEtitleabstractindextext{%
\begin{abstract}

    The structural and spatial arrangements of cells within tissues represent their functional states, making graph-based learning highly suitable for histopathology image analysis. Existing methods often rely on fixed graphs with predefined edges, limiting their ability to capture the true biological complexity of tissue interactions. In this work, we propose ABiG-Net (\underline{A}pproximate \underline{Bi}level Optimization for \underline{G}raph Structure Learning via Neural \underline{Net}works), a novel framework designed to learn optimal interactions between patches within whole slide images (WSI) or large regions of interest (ROI) while simultaneously learning discriminative node embeddings for the downstream image classification task. Our approach hierarchically models the tissue architecture at local and global scales. At the local scale, we construct patch-level graphs from cellular orientation within each patch and extract features to quantify local structures. At the global scale, we learn an image-level graph that captures sparse, biologically meaningful connections between patches through a first-order approximate bilevel optimization strategy. The learned global graph is optimized in response to classification performance, capturing the long-range contextual dependencies across the image. By unifying local structural information with global contextual relationships, ABiG-Net enhances interpretability and downstream performance. Experiments on two histopathology datasets demonstrate its effectiveness: on the Extended CRC dataset, ABiG-Net achieves $97.33\pm1.15\%$ accuracy for three-class colorectal cancer grading and $98.33\pm0.58\%$ for binary classification; on the melanoma dataset, it attains $96.27\pm0.74\%$ for tumor–lymphocyte ROI classification.

\end{abstract}
}

\clearpage
\onecolumn
\thispagestyle{empty}
\begin{center}
\vspace*{\fill}
\large
This work has been submitted to the IEEE for possible publication.\\
Copyright may be transferred without notice, after which this version\\
may no longer be accessible.
\vspace*{\fill}
\end{center}
\newpage
\twocolumn
\setcounter{page}{1}

\maketitle
\IEEEdisplaynontitleabstractindextext  % prints the above block

\subsection{Abbreviations and Acronyms}
The abbreviations and acronyms used in this paper
are listed in the Table~\ref{tab:table1}.

\section{Introduction}
\label{sec:introduction}

Histopathology image analysis has emerged as a critical component of digital pathology, playing a central role in disease diagnosis, prognosis estimation, and therapeutic monitoring. Advances in deep learning, particularly convolutional neural networks (CNNs)~\cite{cnn1_review, cnn2_review, cnn3_review, cnn_review4, cnn_review_11, cnn_review_6}, have significantly improved diagnostic accuracy compared to traditional methods. A common strategy in these approaches is to divide whole slide images (WSIs) or large regions of interest (ROIs) into smaller patches, process them individually during training and testing, and then aggregate patch-level predictions into image-level outcomes. Although this strategy is effective, its reliance on smaller patch sizes creates a key limitation in capturing broader spatial contexts or long-range dependencies. Furthermore, this strategy can incorporate irrelevant background regions, which restricts the model's learning capacity.

\begin{table}[htbp]
\centering
\caption{Abbreviations and acronyms used in this paper.}
\label{tab:table1}
\begin{tabular}{ll}
\hline
\textbf{ABiG-Net} & Approximate Bilevel Optimization for Graph \\ & Structure Learning via Neural Networks \\
\textbf{WSI}      & Whole Slide Image \\
\textbf{ROI}      & Region of Interest \\
\textbf{CRC}      & Colorectal Cancer \\
\textbf{CG}       & Cell Graph \\
\textbf{VD}       & Voronoi Diagram \\
\textbf{DT}       & Delaunay Triangulation \\
\textbf{MST}      & Minimum Spanning Tree \\
\textbf{NN}       & Nearest Neighbors \\
\textbf{GCN}      & Graph Convolutional Network \\
\textbf{GNN}      & Graph Neural Network \\
\textbf{GAT}      & Graph Attention Network \\
\textbf{CNN}      & Convolutional Neural Network \\
\textbf{JK}       & Jumping Knowledge \\
%\textbf{ViT}      & Vision Transformer \\
\textbf{MLP}      & Multilayer Perceptron \\
\textbf{GLCM}     & Gray Level Co-Occurrence Matrix \\
\textbf{ReLU}     & Rectified Linear Unit \\
\textbf{BLO}      & Bilevel Optimization \\
\textbf{FOA}      & First Order Approximate \\
\textbf{SOA}      & Second Order Approximate \\
%\textbf{C2P-GCN}  & Cell-to-Patch Graph Convolutional Network\\
%\textbf{CGC-Net}  & Cell Graph Convolutional Network\\
%\textbf{HAT-Net}  & Hierarchical Transformer Graph Neural Network\\
\hline
\end{tabular}
\end{table}

To overcome these limitations, graph convolutional network (GCN)-based approaches~\cite{gcn_ref1, gcn_ref2, gcn_ref3, gcn_ref4, gcn_ref5, gcn_ref6, gcnreview} that leverage cell-graph representations~\cite{cg} have gained increasing attention in histopathology image analysis. By explicitly modeling spatial and topological relationships among cells, these methods capture the complex organizational patterns of WSIs and large ROIs, thereby enhancing both interpretability and diagnostic accuracy. However, most existing GCN-based approaches rely on fixed input graphs with predefined edge connections, which can introduce strong inductive biases and may fail to reflect the true underlying biological tissue interactions. In addition, such rigid graph constructions often lack adaptability between different types of tissue, staining variations, and disease contexts, limiting their generalization. This highlights the need for more flexible and data-driven graph structure learning frameworks that can dynamically infer meaningful connectivity patterns while preserving biological relevance.

To reduce reliance on heuristic-based connectivity, adaptive graph structure learning for WSI analysis has recently gained significant attention. These methods mainly involve applying a learned transformation to dynamically update the adjacency matrix based on model gradients \cite{adnan, liu, shu, li}. Alternatively, some approaches construct graphs using learned CNN filters where the parameters are updated dynamically according to the model gradients \cite{ding2023}. Some methods also use patch-based graph structure learning, which dynamically selects patches as nodes on each slide during training \cite{kim, BEHZADI}. Despite these promising advances, most existing methods jointly optimize graph structures and node embeddings in a single-level optimization process. Such unified optimization may constrain the flexibility of the graph structure learning process, potentially limiting the exploration of meaningful connectivity patterns.

To alleviate this issue, in this work, we introduce ABiG-Net, \underline{A}pproximate \underline{Bi}level Optimization for \underline{G}raph Structure Learning via Neural \underline{Net}works, a novel bilevel optimization-based framework for histopathology image analysis. ABiG-Net is designed to simultaneously learn optimal adjacency structures, which capture interactions among image patches within WSIs or large ROIs, and their corresponding node embeddings, thereby enhancing downstream image classification performance. 

ABiG-Net involves a two-stage hierarchical graph construction process: Initially, we divide the WSIs or large ROIs into overlapping patches. Within each patch, we construct patch-level graphs, where individual cells serve as nodes, and the edges represent proximity- and similarity-based connections among these cells. We then extract meaningful features from each patch to quantify local tissue structures effectively. Next, we construct a global, image-level graph where each node corresponds to an individual patch, and the node attribute represents the respective patch-level graph feature. To effectively learn optimal connectivity between these patches, we leverage a first-order approximate bilevel optimization approach \cite{darts} employing two distinct neural network modules: (i) a parametric adjacency generation network ($\Phi_{\psi}$), parameterized by $\psi$, which explicitly learns connectivity structures between patches within a WSI or a large ROI; and (ii) a multilayer GCN-based network $f_{\theta}$, parameterized by $\theta$, which learns patch-level embeddings and performs image classification based on the learned adjacency structure. Our parametric adjacency matrix employs the Gumbel-Sigmoid reparameterization trick \cite{gumbel, gumbel2}, which enables a differentiable sampling of discrete adjacency structures by adding Gumbel noise and then applying a temperature-controlled sigmoid activation. This mechanism explicitly samples sparse, binary adjacency structures in a differentiable and stochastic manner, enabling end-to-end gradient-based optimization. The framework optimizes both neural networks jointly through two nested optimization processes. At the lower level, we optimize the embedding parameters $\theta$ to minimize the training classification loss with fixed adjacency parameters $\psi$. At the upper level, we optimize the adjacency parameters $\psi$ by minimizing the validation loss. The trained parameters, once finalized, are used to generate adjacency matrices and classify unseen images. The key contributions of this work are as follows:

\begin{itemize}

\item To the best of our knowledge, this is the first work to employ a bilevel optimization framework to jointly learn the image-level graph structure and the classification model for WSIs or large ROIs, thereby significantly enhancing downstream performance.

\item We design a dual-stage graph construction in which patch-level cell graphs capture local tissue micro-architecture, while an image-level graph models long-range context among different regions (patches) of the WSI/large ROI.

\item We introduce the Gumbel–Sigmoid reparameterization trick into our adjacency generation mechanism, which not only ensures differentiable and stochastic adjacency sampling but also enables broader exploration of the adjacency space during the early training steps, significantly reducing the risk of convergence to local minima.

\item We demonstrate the visual interpretability of our framework by highlighting key regional interactions among patches, offering insight into the model's decision-making process.

\item We conduct comprehensive experiments on two histopathology datasets— the Extended CRC dataset \cite{extend} and the Melanoma dataset \cite{melanoma_nyu} —demonstrating that ABiG-Net achieves performance on par with or exceeding state-of-the-art CNN- and GCN-based methods.

\end{itemize}

The rest of this paper is structured as follows. Section II provides an overview of the related works. Section III describes the proposed methodology in detail, including the overall framework, graph construction strategy, and learning architecture. Section IV presents the experimental setup and performance evaluation on the datasets used in our work. Section V reports ablation studies conducted to assess the contribution of different components of the model. Finally, Section VI describes the conclusion of the paper with a summary of key findings and outlines potential directions for future research.

\section{Related Work}

In this section, we review the literature relevant to our work. We begin by discussing the foundational concept of cell graphs in digital pathology and their application with traditional machine learning. We then survey more recent approaches that pair cell graphs with modern GNN-based architectures. Finally, we discuss adaptive graph structure learning methods that dynamically infer meaningful connectivity in histopathology images.

\subsection{Cell Graphs in Digital Pathology}

Cell graphs have emerged as an effective tool for modeling the structural organization of cells and the surrounding tissue microenvironment by leveraging graph-theoretic representations. In a cell graph, nuclei or cells are considered as vertices (node) of the graph, and the potential interactions between them are represented as edges \cite{cg, cg_sch, cg_demir}. The underlying assumption is that spatially closest cells are more prone to interact, which allows the implementation of the graph construction strategies such as Delaunay triangulation or the k-nearest-neighbor (kNN). Another approach for modeling cell interactions is the Waxman model, which defines the edge probability between two cells as an exponential decay of their Euclidean distance \cite{cg_waxman, cg_waxman2}. Beyond these classical graph constructions, the authors in \cite{cg_bilgin} presented the ECM-aware cell graph for the modeling and classification of bone tissue, where extracellular matrix information was incorporated by encoding color based features and assigning a color label to each node. This representation demonstrated the importance of incorporating contextual cues from the tissue environment. Once the cell graph is constructed, local cellular features are aggregated into global features, which are then used to train machine learning models, including SVMs, Bayesian classifiers, random forest, decision tree, and kNN \cite{cg}.

\subsection{Cell Graphs with Graph Neural network}

Recently, approaches combining cell graphs with GNN-based frameworks have gained significant popularity in digital pathology. The ability of the GNN models to understand non-trivial topological relationships in cell graphs makes them a powerful tool for disease diagnosis and prognosis. For example, CGC-Net~\cite{cgcnet} was among the first to integrate conventional cell graphs with the GCN-based framework for the colorectal cancer (CRC) classification task. To enhance node representations with valuable contextual information, the model incorporated an Adaptive GraphSAGE \cite{graphsage} module to fuse multi-scale features in a data-driven way. Authors in \cite{gcn_jk} introduced a GCN-based framework enhanced with Jumping Knowledge and GraphSAGE to classify intestinal glands as normal or dysplastic. In addition, the authors benchmarked different message-passing architectures against a classical baseline that combined approximate graph edit distance with a KNN classifier. In \cite{hatnet}, the authors proposed HAT-Net, a hierarchical network for colorectal cancer grading task, which integrates a Graph Isomorphism Network (GIN) \cite{GIN} module with MinCut Pooling \cite{mincutpool} to enhance graph discriminability, while a Transformer module \cite{transformer} was introduced to capture long-range dependencies. The authors proposed HACT-Net in \cite{HACT}, a hierarchical cell-to-tissue graph neural network that jointly models cell-level morphology and tissue-level structures through a multi-scale graph representation. Evaluated on the BRACS breast carcinoma dataset, HACT-Net outperformed CNN and GNN baselines, demonstrating superior performance in multi-class breast cancer subtyping. Another approach is C2P-GCN~\cite{c2pgcn}, in which the authors introduced a dual-stage graph construction framework that integrates local patch-level and global image-level tissue structures. Evaluated on two colorectal cancer datasets from different institutes and demographics, the method achieved comparable or superior performance compared to the state-of-the-art approaches. In \cite{cgsig}, the authors proposed CGSignature, a cell-graph based GNN framework that integrates spatial and morphological features from multiplexed immunohistochemistry images of gastric cancer. By benchmarking four GNN variants (GCNSag, GCNTopK, GINSag, and GINTopK), they showed that CGSignature significantly outperforms the TNM staging system in stratifying patient survival. The authors proposed CG-JKNN in \cite{vasu}, a cell graph neural network with a jumping knowledge mechanism, to classify acid-fast bacilli and activated macrophages from lung tissue WSIs of Diversity Outbred mice. By integrating biologically informed graph construction with morphology and graph features, the method achieved strong performance, closely aligning with pathologists criteria for the diagnosis of TB. Collectively, these advances show the potential of combining cell graphs and GCN frameworks to advance digital pathology.

\subsection{Graph Structure Learning in Histopathology Images}

In histopathology, most GNN-based methods employ predefined graphs where the edge structure is fixed in advance. Histological images provide no explicit rule for deciding whether a connection between two nodes should exist, which leads to a diversity of graph construction heuristics. These graph constructions are rarely guided by biological knowledge, which can impose inductive biases that deviate from the true organization of the tissue. To overcome these heuristic graph structures in GNN-based histopathology analysis, adaptive graph structure learning approaches have recently gained popularity.

For instance, in \cite{adnan}, authors proposed dynamically learning graph connectivity for lung cancer subtype classification by representing WSIs as fully connected graphs of representative patches, integrating global image context and local pairwise relationships. Authors in~\cite{hou} presented a spatial-hierarchical GNN with a dynamic structure learning module that jointly learns node embeddings and edge connectivity during training, employing k-NN with distance constraints. In \cite{liu}, the authors introduced a sparse adaptive graph learning mechanism for survival prediction, employing cosine similarity-based adjacency matrices that evolve adaptively during model training. Authors in \cite{li} constructed WSI graphs using distinct ``head" and ``tail" embeddings for patches, dynamically forming edges based on pairwise similarities refined via a knowledge-aware attention mechanism. Slide-level graphs were generated in \cite{shu} by dynamically modeling inter-slide correlations, enhancing performance through a combination of multiple instance learning (MIL) and GNN modules. In \cite{kim}, the authors developed MicroMIL, a graph-based MIL framework using a deep clustering approach and Gumbel-Softmax sampling to dynamically select representative microscopy images as nodes, establishing connections based on feature similarity. Authors in \cite{BEHZADI} introduced a framework that selects a discriminative subset of patches for each WSI using an autoencoder to extract latent patch features and a multi-head attention mechanism to score their relevance. The selected patches are then connected via k-NN to construct a compact patch graph for downstream GNN-based analysis. Authors in \cite{ding2023} proposed a CNN-filter based method where feature maps generated by a CNN are treated as node embeddings for graph construction. The resulting nodes are connected using k-NN, enabling the learned graphs to capture spatial dependencies between tissue regions for downstream GNN analysis. In~\cite{liu2024}, authors extended the CNN-based graph construction methodology by hierarchically applying convolutional layers, building graphs from learned feature maps, and incorporating message passing with a transformer to capture long-range dependencies. At each hierarchical layer, they employed a dilated k-NN strategy, which expands each node’s receptive field and improves feature aggregation across scales.

Despite promising results, most existing methods learn graph structures and perform downstream tasks through a single-level optimization, limiting structure flexibility and biologically meaningful connectivity. To address this, we propose a bilevel optimization framework that decouples adjacency structure learning from node embedding optimization. This approach reduces heuristic biases and improves interpretability and classification performance in histology image analysis.

\section{Methodology}

We begin by dividing each WSI or ROI into appropriately sized overlapping or nonoverlapping patches. Within each patch, we construct patch-level graphs in which we detect nuclei locations along with morphological, intensity, and texture-based features. These form the basis for constructing local cell graphs on each patch that model how neighboring cells interact. At the global scale, the patches themselves are treated as nodes in an image-level graph. Instead of relying on a fixed adjacency structure, we employ a parametric adjacency generator that adaptively learns connections between patches under a bilevel optimization scheme. To ensure clarity and consistency, all mathematical notations and symbols used to describe the proposed methodology, including graph construction, feature representation, and optimization variables, are summarized in Table~\ref{tab:notation}.

\begin{table}[h]
\centering
\caption{Notation used throughout the paper}
\label{tab:notation}
\renewcommand{\arraystretch}{1.2}
\setlength{\tabcolsep}{4pt}
\begin{tabular}{|p{0.7cm}|p{2.8cm}|p{0.9cm}|p{3cm}|}
\hline
$g_{cg}$ & Patch-level cell graph & $V_p, E_{cg}$ & Set of nodes ($V_p$) and edges ($E_{cg}$) in cell graph \\
\hline
$A_p$ & Patch-level adjacency matrix & $D_p$ & Euclidean distance between cells \\
\hline
$P_m$ & $m$th Voronoi polygon & $x_{\mathrm{vor}}$ & Voronoi features \\
\hline
$x_{\mathrm{del}}$ & Delaunay triangulation features & $x_{\mathrm{mst}}$ & Minimum-spanning-tree (MST) features \\
\hline
$x_{\mathrm{nn}}$ & Nuclei nearest-neighbor features & $x_m$ & Patch-level feature vector (all features) \\
\hline
$X_i$ & Set of patch-level features for image $i$ & $G_I(\psi)$ & Image-level graph parameterized by $\psi$ \\
\hline
$V_I$ & Set of image-level nodes (patches) & $\mathcal{E}_I$ & Set of image-level edges \\
\hline
$A_I(\psi)$ & Image-level adjacency matrix & $f_\theta$ & Image classifier (GCN +MLP) with parameters $\theta$ \\
\hline
$\mathcal{D}_{\text{train}}$ & Training dataset & $\mathcal{D}_{\text{val}}$ & Validation dataset \\
\hline
$\mathcal{D}_{\text{test}}$ & Test dataset & $y_i$ & Label of the $i$th image \\
\hline
$X\!\in\!\mathbb{R}^{n\times d}$ & Patch feature matrix (n patches, d features) & $n, d$ & \# patches; feature dimension \\
\hline
$\Phi_{\psi}$ & Parametric adjacency generator (MLP) & $\psi$ & Parameters of adjacency generator \\
\hline
$g_{s,t}$ & Gumbel noise for edge $(s,t)$ & $\tau$ & Temperature in Gumbel--Sigmoid \\
\hline
$\sigma(\cdot)$ & Sigmoid function & $R(\psi)$ & Sparsity regularizer on $A_I$ ($\ell_1$) \\
\hline
$\lambda_{\text{sparse}}$ & Coefficient for $R(\psi)$ & $d_p$ & Cell-neighborhood radius for $A_p$ edges \\
\hline
$H^{(l)}$ & Node-feature matrix at GCN layer $l$ & $W^{(l)}$ & Weight matrix at layer $l$ \\
\hline
$P^{(l)}$ & Global mean pooled vector of layer $l$ & $P_{\text{concat}}$ & Concatenated pooled features across layers \\
\hline
$z^{(t)}$ & Hidden vector in MLP head & $\hat{Y}$ & Predicted class-probability vector \\
\hline
$L_{\text{train}}$ & Training cross-entropy loss & $L_{\text{val}}$ & Validation loss (with $R(\psi)$) \\
\hline
\end{tabular}
\end{table}

\subsection{Cell-identification} 

\begin{figure}[h]
\centering
\includegraphics[width=6.3cm, height=3.3cm]{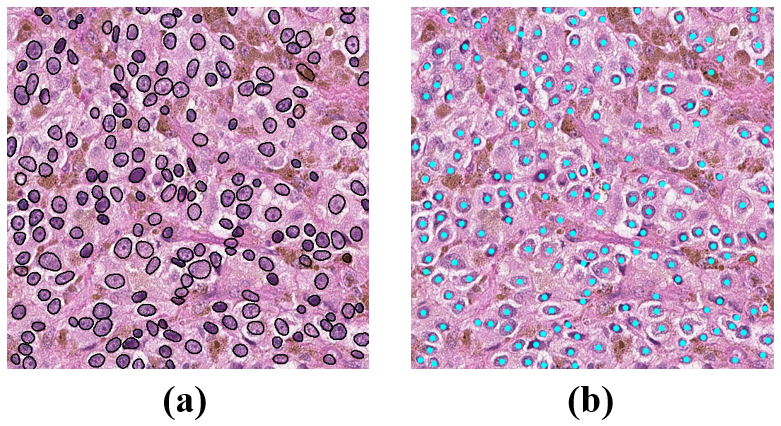}
\caption{Illustration of nuclei detection on a representative histopathology image patch. (a) Segmented nuclei overlaid with contours outlining nuclear boundaries. (b) Centroids (seeds) identified within each segmented region, serving as the basis for patch-level graph construction.}
\label{fig:nuclei_detection}
\end{figure}

After dividing each WSI or large ROI into appropriately sized patches, the next step is to identify individual cells within each patch. Accurate segmentation of nuclei is crucial, as it forms the basis for reliable feature extraction and subsequent graph construction. To achieve this, we employ the pre-trained StarDist 2D model \cite{stardist}, which is specifically designed for nucleus segmentation. StarDist represents each detected object as a star-convex polygon, an approach that effectively handles crowded and overlapping nuclei found in histopathology images. Its ability to generalize across tissue types makes it well-suited for large-scale studies like ours. An illustrative example of nucleus detection is shown in Fig.~\ref{fig:nuclei_detection}.

Once the nuclei are segmented, we extract a set of features from each identified cell. In this work, we compute $12$ hand-crafted features that capture morphological properties (e.g., size, shape, and boundary characteristics), intensity-based measures, and texture attributes from each nucleus. As shown in Table~\ref{tab:cellfeat}, these features capture key aspects of cellular appearance and are later used as node attributes in the constructed graph.

\begin{table}[htbp]
\centering
\caption{Feature definitions used for nuclei characterization.}
\label{tab:cellfeat}
\renewcommand{\arraystretch}{1.2}
\begin{tabular}{|p{0.23\linewidth}|p{0.62\linewidth}|}
\hline
\textbf{Features} & \textbf{Description} \\
\hline
Area & Measures the number of pixels occupied by a nucleus, representing its size. \\
\hline
Major axis length & Length of the ellipse’s major axis that approximates the nucleus shape, indicating its longest dimension. \\
\hline
Minor axis length & Length of the ellipse’s minor axis that approximates the nucleus shape, indicating its shortest dimension. \\
\hline
Eccentricity & Ratio of the distance between the foci of the ellipse fitted to the nucleus and the length of its major axis; reflects how elongated the nucleus is. \\
\hline
Perimeter & Computes the boundary length of the nucleus. \\
\hline
Diameter & The average of the major and minor axes of the ellipse that best fits the nucleus. \\
\hline
Mean intensity & Average pixel intensity within the nucleus, reflecting its overall staining level.\\
\hline
Contrast & Quantifies the variation in gray-level intensity within the nucleus, capturing textural irregularities. \\
\hline
Energy & Sum of squared elements in the gray-level co-occurrence matrix (GLCM) of the nucleus, measuring uniformity of its intensity distribution. \\
\hline
Correlation & Calculate the likelihood of pixel intensity pairs occurring within the nucleus, reflecting internal structural organization. \\
\hline
Homogeneity & Measures how close the GLCM elements of the nucleus are to its diagonal, indicating smoothness of intensity distribution. \\
\hline
ASM Value & Angular second moment; another measure of uniformity in the intensity distribution of the nucleus. \\
\hline
\end{tabular}
\end{table}

\subsection{Patch-level Graph} 

\begin{figure}[h]
\centering
\includegraphics[width=4.5cm, height=6.5cm]{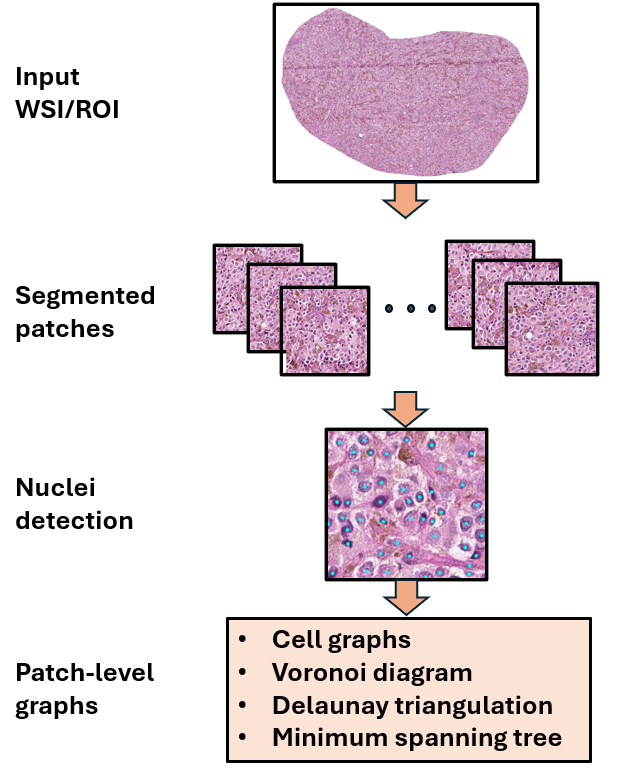}
\caption{An overview of the patch-level graph construction pipeline. An input WSI or large ROI is divided into patches, from which nuclei are detected to generate a set of complementary graphs that capture local patch-level tissue structures.}
\label{fig:patch_level_graph}
\end{figure}

After nuclei detection and extraction of cell-level features, each patch is represented by a set of complementary graphs that capture local cellular interactions and coarser spatial organization. First, an \(\epsilon\)-neighborhood cell graph is constructed on nuclear centroids to model cell-to-cell interactions. To summarize larger-scale cellular arrangement inside a patch, three geometry-driven structures are derived from the same centroids: the Voronoi diagram, its geometric dual- the Delaunay triangulation, and the minimum spanning tree (MST). From these graphs, we compute various hand-crafted features that provide a comprehensive quantitative description of the patch-level topology. An overview of the patch-level graph is presented in Fig.~\ref{fig:patch_level_graph}.

\subsubsection{Cell graph}

\begin{figure}[h]
\centering
\includegraphics[width=7cm, height=7cm]{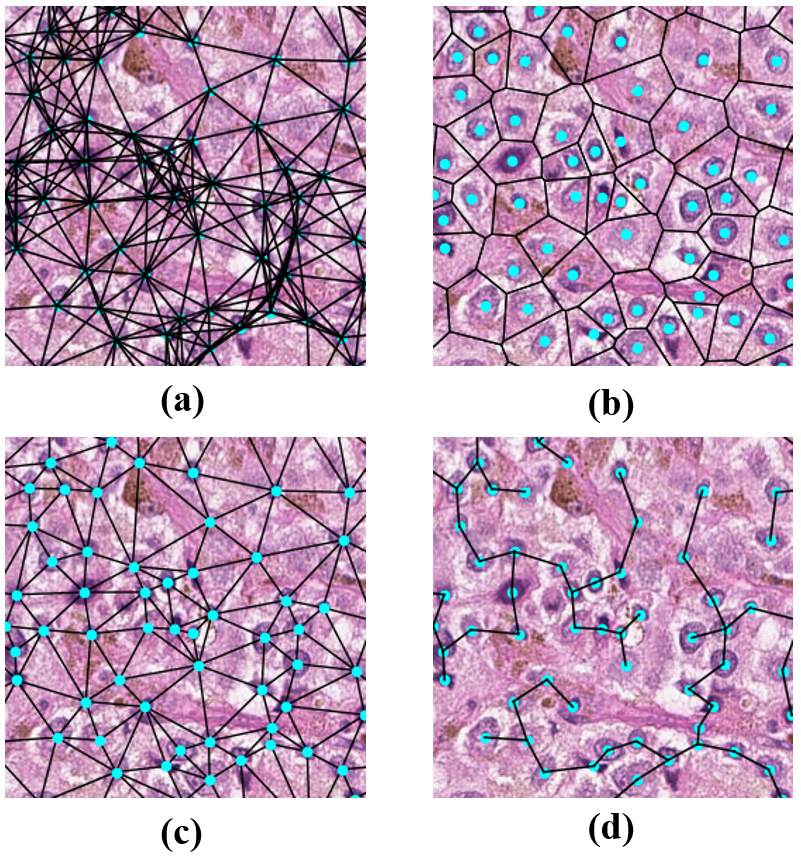}
\caption{Examples of patch-level graph representations. (a) Cell graph constructed using proximity and feature similarity. (b) Voronoi diagram. (c) Delaunay triangulation. (d) Minimum spanning tree. All four graphs are generated from the same set of nuclei centroids within a single representative image patch.}
\label{fig:cell_graphs}
\end{figure}

We represent each patch as a cell graph $g_{cg} = (V_p, E_{cg})$, where the set of nodes $V_p$ corresponds to the segmented cells and the edges $E_{cg}$ encode potential interactions between them. Direct modeling of the true biological interactions is inherently challenging, as such interactions often depend on complex factors that are not directly observable. To approximate these interactions in a computationally feasible and biologically meaningful manner, we incorporate both spatial proximity and phenotypic similarity into the edge definition. Specifically, two cells $u$ and $v$ are connected if (i) the Euclidean distance between their centroids, $D_p(u, v)$, falls below a fixed threshold $d_p$, and (ii) the cosine similarity between their feature vectors $x_u$ and $x_v$ exceeds a threshold $\theta_{sim}$. This formulation can be regarded as a variant of an $\epsilon$ neighborhood graph that also incorporates feature similarity, allowing the graph to capture both local spatial cellular organization and functional similarity. The resulting adjacency matrix $A_p$ is therefore defined as
\begin{equation}
A_p(u, v) =
\begin{cases}
    1, & \text{if } D_p(u, v) < d_p \text{ and } \text{cos}(x_u, x_v) > \theta_{sim}, \\
    0, & \text{otherwise.}
\end{cases}
\end{equation}
An illustrative example of this patch-level cell graph is shown in Fig.~\ref{fig:cell_graphs}(a). This graph representation is designed to highlight local cellular arrangements, such as clustering, crowding, or isolation, which may carry diagnostic or prognostic significance. To further characterize these spatial patterns, we compute a set of $18$ quantitative features $x_{cg}$ from each graph \cite{cg, c2pgcn, flock}, as summarized in Table~\ref{tab:cell_graph_feats}. These features provide a compact description of patch-level organization and serve as a bridge between raw image data and higher-level graph-based learning.

\begin{table*}[!h]
\caption{Cell-graph features}
\label{tab:cell_graph_feats}
\centering
\renewcommand{\arraystretch}{1.15}
\setlength{\tabcolsep}{5pt}
\newcolumntype{P}[1]{>{\raggedright\arraybackslash}p{#1}}
\begin{tabular}{|P{2.8cm}|c|P{0.9cm}|P{12.2cm}|}
\hline
\textbf{Feature type} & \textbf{No.} & \textbf{Feature ID} & \textbf{Description} \\ \hline

\multirow{4}{=}{Connectedness and cliquishness measures}
& \multirow{4}{*}{4} & {01} & Clustering coefficient: Fraction of actual links between a cell's neighbors out of all possible links \\
& & {02} & Average degree: Average number of connections per cell in the graph \\
& & {03} & Number of connected components: Count of isolated parts of the cell graph with no links between them \\
& & {04} & Giant connected component ratio: Fraction of nodes that belong to the largest connected component \\ \hline

\multirow{8}{=}{Distance-based measures}
& \multirow{8}{*}{8} & {05} & Number of vertices: Number of cells in a patch \\
& & {06} & Number of edges: Number of hypothesized communications between cells \\
& & {07} & Average eccentricity: Average of the maximum shortest-path distances (eccentricities) of all nodes \\
& & {08} & Radius: Minimum eccentricity among all cells \\
& & {09} & Diameter: Maximum eccentricity among all cells \\
& & {10} & Number of central points: Number of cells whose eccentricity equals the graph radius \\
& & {11} & Percent of central points: Percentage of cells whose eccentricity equals the graph radius \\
& & {12} & Closeness (average): Inverse of the average shortest-path distance from a cell to all other cells \\ \hline

\multirow{6}{=}{Spectral measures}
& \multirow{6}{*}{6} & {13} & Energy of Laplacian: Sum of deviations of Laplacian eigenvalues from the average degree; reflects irregularity of connectivity \\
& & {14} & Trace of Laplacian: Sum of node degrees (equals twice the number of edges) \\
& & {15} & Upper slope: Slope of normalized Laplacian eigenvalues in $[1\text{--}2]$, reflecting global structural irregularity \\
& & {16} & Lower slope: Slope of normalized Laplacian eigenvalues in $[0\text{--}1]$, reflecting local cell connectivity \\
& & {17} & Largest adjacency eigenvalue: Spectral radius of the adjacency matrix; overall strength of cell connectivity \\
& & {18} & Energy of adjacency: Sum of absolute adjacency eigenvalues; overall strength of cell-cell connectivity \\ \hline

\end{tabular}
\end{table*}

\subsubsection{Voronoi diagram}

The Voronoi diagram partitions the image patch into a set of nonoverlapping polygons $P_1, P_2, ..., P_m$ (disjoint up to shared boundaries), each surrounding its corresponding centroid of the nuclei $v_i \in V_p$. An example of a Voronoi diagram is presented in Fig.~\ref{fig:cell_graphs}(b). Each pixel in the patch is assigned to its nearest centroid, forming the corresponding polygon. The boundaries between these polygons, shown in black, are geometrically defined as the set of points equidistant from the two nearest centroids. From the polygons $\{P_i\}_{i=1}^m$, we compute the total area of the polygons, as well as the mean, standard deviation, minimum/maximum ratio, and disorder for the area, perimeter length, and chord length of the polygon. This yields a set of $13$ Voronoi features $x_{vor}$ \cite{ patchfeats, patchfeats2, flock} for each image patch, as shown in Table~\ref{tab:patch_graph_feats}.

\begin{table}[h]
\caption{Patch-level global cell-graph features}
\label{tab:patch_graph_feats}
\centering
\renewcommand{\arraystretch}{1.15}
\setlength{\tabcolsep}{2pt}
\newcolumntype{P}[1]{>{\raggedright\arraybackslash}p{#1}}
\begin{tabular}{|P{1.7cm}|c|P{0.85cm}|P{5.0cm}|}
\hline
\textbf{Feature type} & \textbf{No.} & \textbf{Feature ID} & \textbf{Description} \\ \hline

\multirow{4}{=}{Voronoi diagram features, $x_{\text{vor}}$}
& 1  & {19}                & Total area of polygons \\
& 4  & {20--23}          & Polygon area: mean, SD, min/max ratio, disorder \\
& 4  & {24--27}          & Polygon chord length: mean, SD, min/max ratio, disorder \\
& 4  & {28--31}        & Polygon perimeter: mean, SD, min/max ratio, disorder \\ \hline

\multirow{2}{=}{Delaunay triangulation features, $x_{\text{del}}$}
& 4  & {32--35}          & Triangle area: mean, SD, min/max ratio, disorder \\
& 4  & {36--39}          & Triangle side length: mean, SD, min/max ratio, disorder \\ \hline

Minimum spanning tree features, $x_{\text{mst}}$
& 4  & {40--43}          & Edge length: mean, SD, min/max ratio, disorder \\ \hline

\multirow{4}{=}{Nearest-neighbor (NN) features, $x_{\text{nn}}$}
& 1  & {44}                 & Density of nuclei in the patch \\
& 1  & {45}                 & Total nuclei count in the patch \\
& 9  & {46--54}           & Distance to $k$-NN ($k{=}3,5,7$): mean, SD, disorder \\
& 15 & {55--69}          & NN count within radius $n$ ($n{=}10,20,\ldots,50$ px): mean, SD, disorder \\ \hline
\end{tabular}
\end{table}

\subsubsection{Delaunay triangulation}

Delaunay triangulation of the vertices $V_p$ connects two nuclei centroids $v_i$ and $v_j$ when their Voronoi polygons $P_i$ and $P_j$ share a common boundary; the resulting Delaunay graph is the geometric dual of the Voronoi diagram. An example is shown in Fig.~\ref{fig:cell_graphs}(c). Each triangle is constructed so that the circle passing through its three vertices does not enclose any other vertex from $V_p$. From these triangles, we compute the area and side length and summarize them by the mean, standard deviation, disorder (std/mean), and minimum/maximum ratio, resulting in $8$ Delaunay-based characteristics, $x_{del}$ per patch \cite{ patchfeats, patchfeats2, flock}, as represented in Table \ref{tab:patch_graph_feats}.

\subsubsection{Minimum spanning tree}

The Minimum Spanning Tree (MST) on the vertex set $V_p$ is the acyclic graph that connects all nuclear centroids using the smallest possible total Euclidean edge length. It has $|V_p|-1$ edges and provides a sparse backbone of spatial organization. Fig.~\ref{fig:cell_graphs}(d) shows a representative example of MST. From the set of MST edge lengths, we compute the mean, standard deviation, disorder (std/mean), and minimum/maximum ratio, yielding $4$ MST‑based features, $x_{mst}$ per patch \cite{ patchfeats, patchfeats2, flock}, shown in Table \ref{tab:patch_graph_feats}.

\subsubsection{Nearest-neighbor descriptors}

Finally, we incorporate a set of nearest-neighbor features that capture cell clustering and density information without explicit graph construction. These features are calculated in two ways. First, to quantify local crowding, we count the number of neighboring nuclei within fixed radii of $10, 20, 30, 40, \text{and } 50$ pixels from each nuclear centroid, and summarize these counts across all nuclei by their mean, standard deviation, and disorder (std/mean), yielding $15$ features. Second, to measure spatial separation, we compute the Euclidean distance from each nucleus to its $k$-nearest neighbor for $k$ values of $3, 5, \text{and } 7$. These distances are similarly summarized across the patch by their mean, standard deviation, and disorder, producing $9$ features. In addition, we extract the total number of nuclei in the patch and the nuclear density. Thus, in total, we extract $26$ nuclei nearest-neighbor based features, $x_{nn}$, from a patch \cite{ patchfeats, patchfeats2, flock} (Table \ref{tab:patch_graph_feats}).

\subsubsection{Feature aggregation}

We construct a comprehensive descriptor for each patch by concatenating features derived from all feature families- cell graphs, Voronoi diagrams, Delaunay triangulations, minimum spanning trees (MSTs), and nearest-neighbor statistics. Specifically, we combine $18$ cell graph features $x_{cg}$, $13$ Voronoi features $x_{vor}$, $8$ Delaunay features $x_{del}$, $4$ MST features $x_{mst}$, and $26$ nearest-neighbor features $x_{nn}$ into a unified vector as
\begin{equation}
x_m = \bigl[\,x_{cg} \,\|\, x_{vor} \,\|\, x_{del} \,\|\, x_{mst} \,\|\, x_{nn}\,\bigr] \in \mathbb{R}^{69}.
\end{equation}
where $\|$ denotes concatenation. For a given WSI or large ROI $i$ with $n_i$ patches, we obtain a set of patch-level feature vectors
\begin{equation}
X_i = \{\,x_m^{(i)}\,\}_{m=1}^{n_i} \in \mathbb{R}^{n_i \times 69}.
\end{equation}
These vectors provide a compact representation of nuclear organization and spatial topology and serve as node attributes in the image-level graph.

\subsection{Image-level Graph}

To explicitly model the global interactions among patches within each WSI or large ROI, we introduce an image-level graph $G_I(\psi) = (V_I, \mathcal{E}_I(\psi))$, where the nodes in the node set $V_I$ correspond to individual patches within a WSI or ROI. The edge set $\mathcal{E}_I(\psi) \subseteq V_I \times V_I$ encodes the connectivity between patches within an image. These connections are explicitly parameterized by a set of trainable parameters $\psi$ represented through a parametric adjacency matrix $A_{I}(\psi)$, whose entries indicate learned connectivity strengths among patches.

\begin{figure*}[t]
\centering
% Adjust the width and height as needed
\includegraphics[width=17cm, height=9cm]{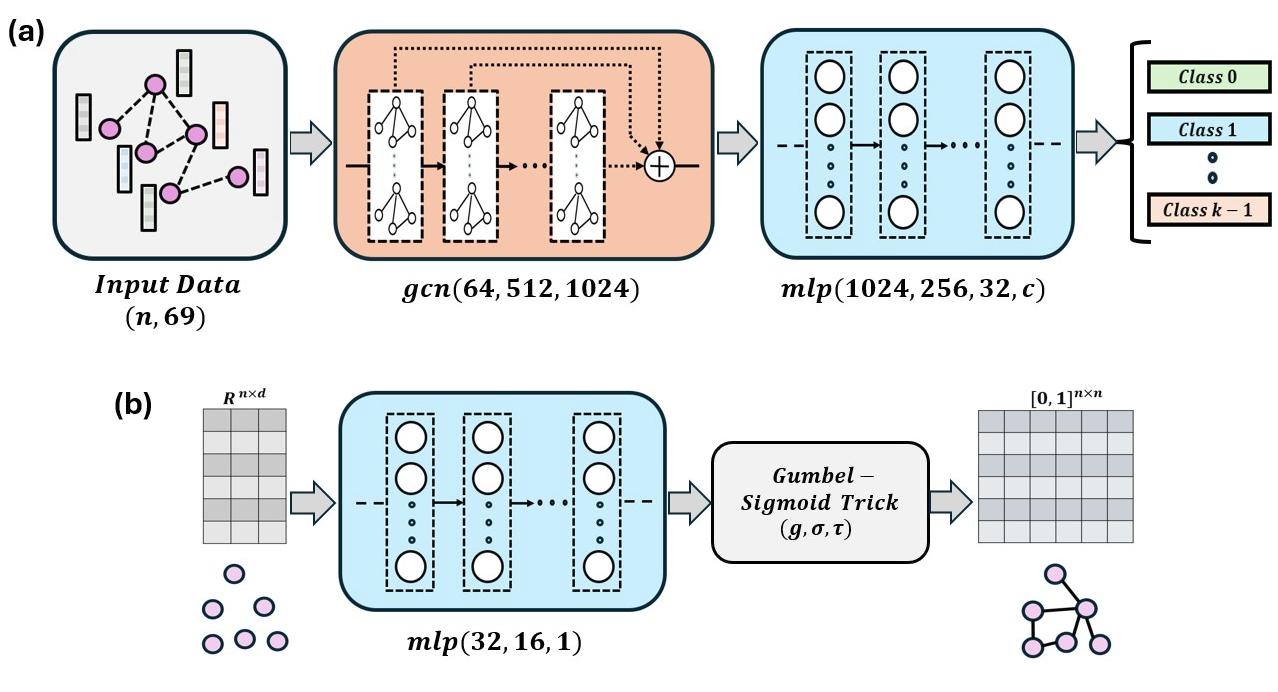}
\caption{Image-level components of the ABiG-Net framework. (a) Image classification network: patch-level feature vectors serve as node features and are processed through a multi-layer GCN with jumping knowledge aggregation. Global pooling and fully connected layers yield image-level predictions. (b) Parametric adjacency generation network: pairwise patch features are processed by an MLP, and adjacency values are sampled via the Gumbel– Sigmoid reparameterization. The resulting sparse, differentiable adjacency matrices define the image-level graph structure used by the GCN.}
\label{fig:abig_net}
\end{figure*}

Now, consider a neural network $f_{\theta}$, with $\theta$ being a set of trainable parameters. It learns discriminative node embeddings for each patch within an image and subsequently aggregates them to form an image-level embedding for classification. Given a training dataset $\mathcal{D}_{\text{train}} = \{(X_i, y_i)\}_{i=1}^{|\mathcal{D}_{\text{train}}|}$ and a separate validation dataset $\mathcal{D}_{\text{val}} = \{(X_j, y_j)\}_{j=1}^{|\mathcal{D}_{\text{val}}|}$, we formulate the joint optimization of the classifier parameter $\theta$ and the adjacency parameter $\psi$ as a bilevel optimization problem. At the lower-level, we fix $\psi$ and optimize the classifier parameters $\theta$ by minimizing the classification loss on the training dataset. At the upper-level, we utilize the optimized $\theta^*(\psi)$ to learn the adjacency parameters $\psi$ by minimizing validation loss combined with a regularization term $\mathcal{R}(\psi)$. Formally,
\begin{equation}
\begin{aligned}
    \psi^* &= \underset{\psi}{\arg\min}\; \mathcal{L}_{\text{val}}\left(\theta^*(\psi), \psi\right) + \mathcal{R}(\psi) \\    
    &= \arg\min_{\psi} \frac{1}{|\mathcal{D}_{\text{val}}|} \sum_{(X, y) \in \mathcal{D}_{\text{val}}} \mathcal{L}(f_{\theta^*(\psi)}(X, A_I(\psi)), y) + \mathcal{R}(\psi)
\end{aligned}
\end{equation}
\begin{equation}
\begin{aligned}
    \text{s.t.} \quad \theta^*(\psi) &= \underset{\theta}{\arg\min}\; \mathcal{L}_{\text{train}}(\theta, \psi) \\    
    &= \arg\min_{\theta} \frac{1}{|\mathcal{D}_{\text{train}}|} \sum_{(X, y) \in \mathcal{D}_{\text{train}}} \mathcal{L}(f_{\theta}(X, A_I(\psi)), y),
\end{aligned}
\end{equation}
where, $\mathcal{L}$ denotes the cross-entropy loss used for classification and $\mathcal{R}(\psi)$ is the regularization term used to encourage sparsity in the learned adjacency matrix. In this work, we apply a first-order approximate bilevel optimization approach \cite{darts}. Hence, when computing gradients to optimize the upper-level parameters $\psi$, the gradient formulation is approximated as:
\begin{equation}
\begin{aligned}
\nabla_{\psi}\left[\mathcal{L}_{\text{val}}(\theta^*(\psi), \psi) + \mathcal{R}(\psi)\right] 
&= {\frac{\partial \mathcal{L}_{\text{val}}}{\partial \psi}} \\
&+ {\frac{\partial \mathcal{L}_{\text{val}}}{\partial \theta^*(\psi)}\frac{\partial \theta^*(\psi)}{\partial \psi}}
+ \nabla_{\psi}\mathcal{R}(\psi)\\
& \approx \frac{\partial \mathcal{L}_{\text{val}}}{\partial \psi} + \nabla_{\psi}\mathcal{R}(\psi),
\end{aligned}
\end{equation}
here, due to the first-order approximation (FOA), we neglect the second order term. We explicitly assume $\partial \theta^*(\psi)/\partial \psi \approx 0.$ Our FOA effectively balances computational efficiency and model performance, enabling direct learning of the graph structure without the overhead associated with second-order methods. A full second-order approach involves computing Hessian matrices, either explicitly or implicitly, which is computationally expensive and memory-intensive. Even efficient second-order approximations (SOA) like Hessian-vector products used in the highly influential Model-Agnostic Meta-Learning (MAML) framework \cite{maml} add considerable complexity. In fact, authors demonstrated on the MiniImagenet dataset that the FOA version achieved nearly the same performance as the full SOA method, while reducing computation time by approximately $33\%$. Inspired by this compelling trade-off, we adopt a first-order approximation for our framework.

In this study, we use a multilayer GCN-based network to represent $f_\theta$. The multilayer GCN will be followed by a series of linear layers and a softmax classification layer. Each node in the graph corresponds to an image patch, and the connections between these patches are determined by the learned adjacency matrix $A_{I}(\psi)$. Node features (patch-level feature vectors) are passed through sequential GCN layers, each followed by a ReLU activation and dropout. For layers $l=0,1,2,...,L-1$, the GCN stack is

\begin{equation}
    H^{(l+1)} = \text{Dropout}\left( \text{ReLU} \left({\text{GCN}}_l\left( H^{(l)}, A_I (\psi)\right) \right) \right),
\end{equation}
where, $H^{(l)} \in \mathbb{R}^{{n \times h_l}}$ is the node feature matrix at layer $l$, and the input to the network is $H^{(0)} = X \in \mathbb{R}^{n \times d}$ is the set of $d$ dimensional feature vector for $n$ patches within a WSI or ROI obtained from patch-level graph. Here, $d = 69$. The learnable weight matrix at layer $l$ is $W^{(l)} \in \mathbb{R}^{h_l \times h_{l+1}}$. We incorporate a concatenation-based ``jumping knowledge" mechanism into our network,  originally introduced in~\cite{jump}. This mechanism enables the network to aggregate information from all intermediate layers rather than relying solely on the final representation, ensuring that information learned at different depths contributes to the final graph embedding. Specifically, we apply global mean pooling to the feature representation of each layer and then concatenate the pooled features from all GCN layers as:

\begin{equation}
\label{eq:concat}
P_{\text{concat}} \;=\; \big[P^{(1)} \;\|\; P^{(2)} \;\|\; \cdots \;\|\; P^{(L)}\big],
\end{equation}
where,
\begin{equation}\nonumber
P^{(l)} = \text{GlobalMeanPool}\left(H^{(l)}\right).
\end{equation}
$P_{\text{concat}}$ is then fed into a sequence of linear layers, each followed by ReLU activation and dropout as

\begin{equation}
z^{(t+1)} = \text{Dropout}\left( \text{ReLU}\left( w^{(t)}_{\text{lin}} z^{(t)} + b^{(t)}_{\text{lin}} \right) \right),
\label{eq:lin_layer}
\end{equation}
where, $z^{(0)} = P_{\text{concat}}$. In~\eqref{eq:lin_layer}, the weights and biases for the linear layers are denoted as $w^{(t)}_{\text{lin}}$ and  $b^{(t)}_{\text{lin}}$ respectively, where the linear layers $t = 0,1,...,N-1$. Finally, a softmax layer is used to produce the class probability distribution $\hat{Y}$ as follows:

\begin{equation}
 \hat{Y} = \text{Softmax}\left( w^{(N)}_{\text{lin}} z^{(N)} + b^{(N)}_{\text{lin}} \right).   
\end{equation}
The multi-layer GCN architecture used to represent $f_\theta$ along with the dimensions of each layer is visualized in Fig. \ref{fig:abig_net} (a).

\subsubsection*{Parametric Adjacency Generator}

To learn the connectivity structure among patches, we define a parametric function $\Phi_\psi (X)$ that maps the patch-level feature matrix $X \in \mathbb{R}^{n \times d}$ to a pairwise logit matrix in $\mathbb{R}^{n \times n}$. Formally,

\begin{equation}
    \Phi_{\psi} (X) : \mathbb{R}^{n \times d} \rightarrow \mathbb{R}^{n \times n},
\end{equation}
where the function $\Phi_{\psi}$ is a neural network with learnable parameters $\psi$. In this work, this function is realized through a multilayer perceptron (MLP) as shown in Fig.~\ref{fig:abig_net} (b). For each pair of nodes $(s,t)$, we concatenate their features into a $2d$-dimensional vector, $\left[x_s \;\|\; x_t\right] \in \mathbb{R}^{2d}$, which is processed individually by MLP to generate a scalar logit representing the connectivity strength:

\begin{equation}
\Phi_{\psi}\left(x_s, x_t\right) = \text{MLP}_{\psi}\left(\left[x_s \;\|\; x_t\right]\right).
\end{equation}
%Here, $\left[x_s \;\|\; x_t\right]$ denotes the concatenation of features from a pair of patches. The function yields a scalar value (logit) showing the strength of connection between two patches. 
Now, to obtain discrete adjacency decisions (binary edges) while preserving differentiability, we utilize the Gumbel-Sigmoid reparameterization trick \cite{gumbel, gumbel2}. Directly sampling binary edges is non-differentiable, which prevents gradient-based learning. In the Gumbel–Sigmoid scheme, independent Gumbel noise is added to the pairwise logits produced by the MLP, and the result is scaled by a temperature parameter and passed through a sigmoid. This yields a continuous approximation to the binary edges that remain differentiable with respect to the logits, thereby enabling end-to-end training. Formally,

\begin{equation}
    A_{I}(\psi)_{s,t} = \sigma\left(\frac{\Phi_{\psi}(x_s, x_t) + g_{s,t}}{\tau}\right), \quad g_{s,t} \sim \text{Gumbel}(0,1)
\end{equation}
where, $g_{s,t}$ is the Gumbel noise, and $\sigma(\cdot)$ is the sigmoid function. The temperature parameter $\tau$ controls the sharpness of the edge distribution and is gradually annealed during training according to

\begin{equation}
     \tau^{(t+1)} = \max\left(\tau_{\text{min}},\, \tau^{(t)} \times \gamma\right). 
\end{equation}
Here, $\tau^{(t)}$ is the temperature at iteration $t$, and,   $\tau^{(0)} = \tau_{\text{init}}$ is the initial temperature, $\tau_{\text{min}}$ is the minimum temperature, and $\gamma$ is the decay rate. 
The Gumbel-Sigmoid transformation produces adjacency values in the range $(0,1)$ that approximate binary edges but remain smooth functions of the logits, noise, and temperature. The introduced stochasticity promotes broader exploration of the candidate graph structures and avoids early convergence to local minima. As training progresses and $\tau$ decreases, the relaxed adjacency values become increasingly sharp, converging towards near-binary decisions while preserving differentiability for gradient-based optimization.

Since the interactions between patches are naturally undirected, we enforce symmetry as

\begin{equation}
A_{I}(\psi) = \dfrac{A_{I}(\psi) + A_{I}(\psi)^{T}}{2}.
\end{equation} 
After enforcing symmetry, we add self-loops to the graph by setting the diagonal elements of the adjacency matrix to $1$, i.e.; $A_{I}(\psi)_{j,j} = 1$. This ensures each node includes its own features during graph convolution operation. To promote sparsity in the learned adjacency matrix and avoid overly dense connections, we incorporate a sparsity regularization term. We apply an L1-norm penalty directly on the adjacency matrix $A_{I}(\psi)$. Formally,

\begin{equation}
\mathcal{R}(\psi) = \lambda_{\text{sparse}}\|A_I(\psi)\|_1 = \lambda_{\text{sparse}}\sum_{j=1}^{n}\sum_{k=1}^{n}|A_I(\psi)_{j,k}|,
\end{equation}
where, $\lambda_{\text{sparse}}$ is a hyperparameter controlling the degree of sparsity on $A_I$. Once we finalize $\theta^*$ and $\psi^*$, for images in our test set we generate the adjacency $A_{test}$. Then, we compute the class probabilities using the trained model as $\hat{Y} = f_{\theta^*}(X_{\text{test}}, A_{\text{test}})$, where $X_{test}$ is the set of patch-level feature vectors of an image in the test set. Finally, we obtain the class label by selecting the class with the highest probability, $\arg\max(\hat{Y})$. The overall learning process is summarized in Algorithm 1.

\begin{algorithm}[h]
\caption{Bilevel Learning of Graph Structure and Image Classifier}
\label{alg:bilevel_gcn}
\begin{algorithmic}[1]
\REQUIRE $\mathcal{D}_{\text{train}}$, $\mathcal{D}_{\text{val}}$, learning rates $\eta_{\theta}, \eta_{\psi}$, iterations $T$, temperature parameters $\tau_{\text{init}}, \tau_{\text{min}}, \gamma$
\ENSURE Trained classifier $\theta^{*}$, trained adjacency generator $\psi^{*}$

\STATE Initialize parameters: $\theta$, $\psi$, $\tau \leftarrow \tau_{\text{init}}$
\FOR{$t = 1$ to $T$}

\STATE \textbf{Upper-level optimization: Generator update (fix $\theta$)}
\STATE Set $f_{\theta}$ \textbf{eval}; $\Phi_{\psi}$ \textbf{train}
\FOR{minibatch $(X_{\text{val}}, Y_{\text{val}}) \sim \mathcal{D}_{\text{val}}$}
    \STATE logits $\leftarrow \Phi_{\psi}(X_{\text{val}})$
    \STATE $A_I \leftarrow \text{GumbelSigmoid}(\text{logits}, g, \sigma, \tau)$
    \STATE $A_I \leftarrow \tfrac{1}{2}(A_I + A_I^{T})$; $\text{diag}(A_I) \leftarrow 1$ \hfill
    %\STATE $\text{diag}(A_I) \leftarrow 1$ \hfill \% add self-loops
    \STATE $\hat{Y} \leftarrow f_{\theta}(X_{\text{val}}, A_I)$
    \STATE $\mathcal{L}_{\text{val}} \leftarrow \text{CrossEntropy}(\hat{Y}, Y_{\text{val}}) + \lambda_{\text{sparse}}\|A_I\|_1$
    \STATE $\psi \leftarrow \psi - \eta_{\psi}\nabla_{\psi}\mathcal{L}_{\text{val}}$
\ENDFOR

\STATE \textbf{Lower-level optimization: Classifier update (fix $\psi$)}
\STATE Set $\Phi_{\psi}$ \textbf{eval}; $f_{\theta}$ \textbf{train}
\FOR{minibatch $(X_{\text{train}}, Y_{\text{train}}) \sim \mathcal{D}_{\text{train}}$}
    \STATE logits $\leftarrow \Phi_{\psi}(X_{\text{train}})$
    \STATE $A_I \leftarrow \text{GumbelSigmoid}(\text{logits}, g, \sigma, \tau)$
    \STATE $A_I \leftarrow \tfrac{1}{2}(A_I + A_I^{T})$; $\text{diag}(A_I) \leftarrow 1$ \hfill
    %\STATE $\text{diag}(A_I) \leftarrow 1$ \hfill \% add self-loops
    \STATE $\hat{Y} \leftarrow f_{\theta}(X_{\text{train}}, A_I)$
    \STATE $\mathcal{L}_{\text{train}} \leftarrow \text{CrossEntropy}(\hat{Y}, Y_{\text{train}})$
    \STATE $\theta \leftarrow \theta - \eta_{\theta}\nabla_{\theta}\mathcal{L}_{\text{train}}$
\ENDFOR

\STATE \textbf{Anneal temperature:} $\tau \leftarrow \max(\tau_{\text{min}}, \tau \times \gamma)$

\ENDFOR
\RETURN $\theta^{*} \leftarrow \theta$, $\psi^{*} \leftarrow \psi$
\end{algorithmic}
\end{algorithm}

\section{Experiments}
\subsection{Dataset}

We evaluated the performance of ABiG-Net on two datasets: Dataset I, the Extended CRC dataset~\cite{extend} from the University of Warwick (UK), and Dataset II, the Melanoma dataset \cite{melanoma_nyu} collected at NYU Langone Health (New York, USA).

\begin{figure}[h]
\centering
\includegraphics[width=8.75cm, height=4cm]{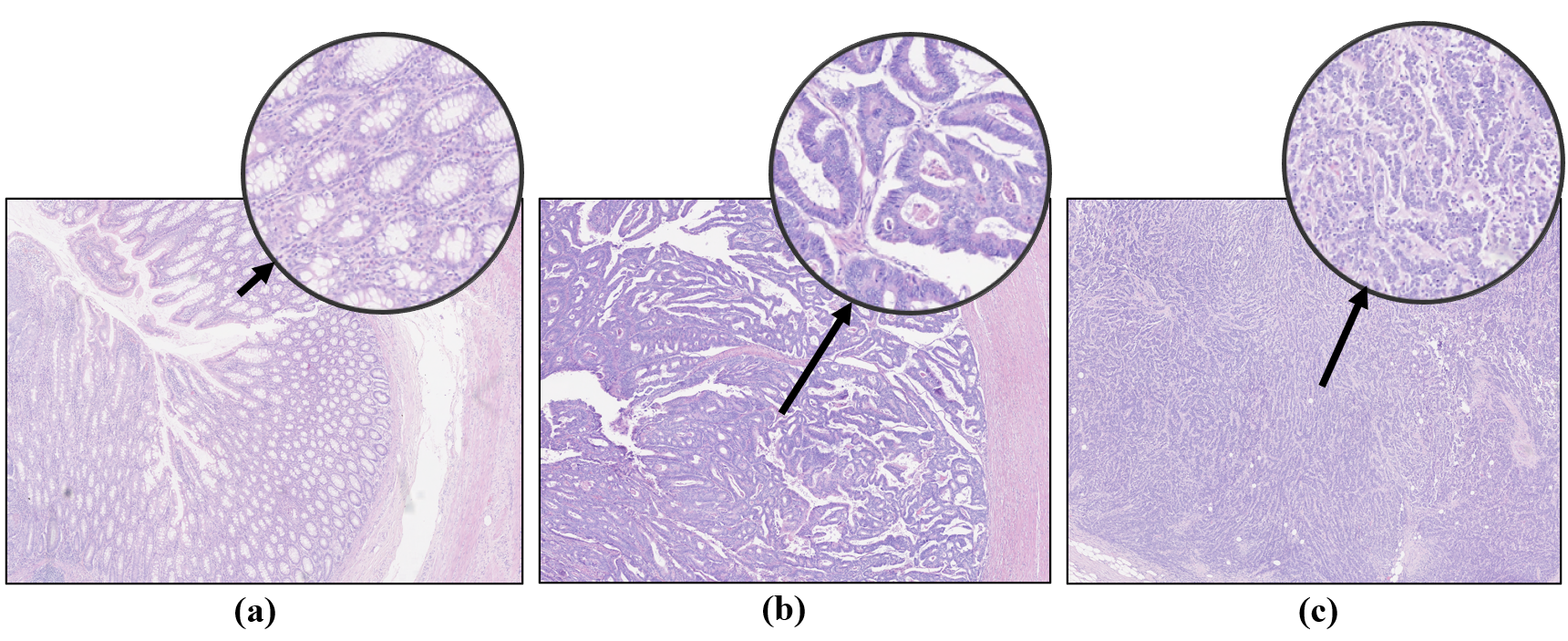}
\caption{Representative visual fields from the Extended CRC dataset. (a) Normal colorectal tissue, (b) Low-grade colorectal cancer, and (c) High-grade colorectal cancer. The circular inset in each panel shows a higher magnification view of the tissue architecture.}
\label{fig:ext_crc_sample}
\end{figure}

Dataset I is derived from $68$ H\&E-stained WSIs of colorectal tissue, scanned at $20\times$ magnification using high-resolution digital scanners. From these WSIs, a total of $300$ representative visual fields were extracted, each capturing distinct regions of the tissue with resolutions of approximately $5000 \times 7300$ or $4548 \times 7520$ pixels. These visual fields capture diverse morphological patterns observed in colorectal cancer. The dataset is categorized into three classes: normal ($120$ images), low-grade cancer ($120$ images), and high-grade cancer ($60$ images). Representative examples from each of the three categories are shown in Fig.~\ref{fig:ext_crc_sample}.

\begin{figure}[h]
\centering
\includegraphics[width=8.75cm, height=5.0cm]{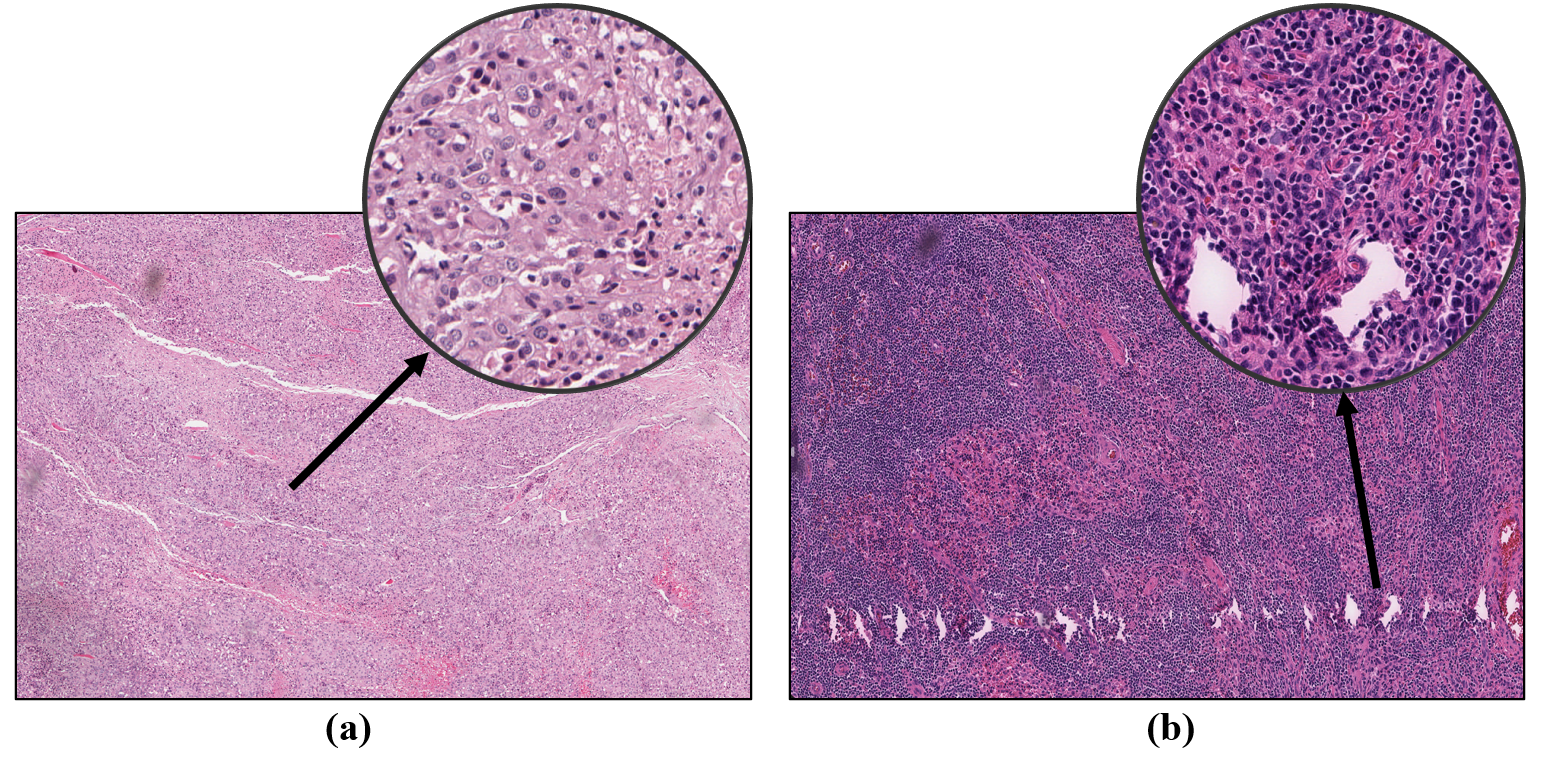}
\caption{Representative visual fields from the Melanoma dataset. (a) Tumor regions, and (b) Lymphocyte-rich regions. Each inset shows a higher-magnification view of the cellular morphology.}
\label{fig:melanoma_data}
\end{figure}

Our Dataset II comprises $153$ H\&E–stained WSIs of metastatic melanoma tissue. The slides were digitized at either $20 \times$ or $40 \times$ magnification using high-resolution whole-slide scanners, with all $40\times$ WSIs subsequently downsampled to $20 \times$ for consistency. From these WSIs, we extracted large regions of interest (ROIs) corresponding to lymphocyte-rich and tumor-rich areas, as annotated by a board-certified pathologist. In total, $1072$ ROIs were obtained, with an average resolution of $4190 \times 4240$ pixels, including $544$ lymphocyte regions and $528$ tumor regions. Representative examples from each category are shown in Fig.~\ref{fig:melanoma_data}.

%Dataset II is a subset of Dataset I, containing $139$ H\&E-stained images from $37$ patients, with an average resolution of $4548 \times 7520$ pixels at $20\times$ magnification, categorized into normal ($71$ images), low-grade ($33$ images), and high-grade cancer ($35$ images). Dataset III consists of $717$ large ROIs, with $355$ cancer and $362$ normal images at $40\times$ magnification, with an average size of $10000\times10000$ pixels. 

\subsection{Implementation}

For the Extended CRC dataset (Dataset I), each large ROI is partitioned into smaller, overlapping patches. We systematically examine multiple configurations by varying patch sizes ($512 \times 512$, $768 \times 768$, and $1024 \times 1024$ pixels) with a stride of $256$ pixels. A patch size of $768 \times 768$ with a stride of $256$ consistently yields the best performance; therefore, this setting is adopted in our experiments. For fair comparison with other methods, we employ a three-fold cross-validation at the image level using the same fold partitions reported in previous works \cite{c2pgcn, extend, hatnet}. In each round, two folds are designated for model development, with $80\%$ of their images are used for training and $20\%$ are used for validation. The remaining fold is reserved exclusively as an independent test set. This procedure is repeated three times so that each fold acts once as the test set, and the reported results reflect the average performance across the three folds.

For the Melanoma dataset (Dataset II), we apply the same patch and stride configuration, exploring patch sizes of $512 \times 512$, $768 \times 768$, and $1024 \times 1024$ pixels with a stride size of $256$ pixels. In this case, a patch size of $512 \times 512$ achieves the most reliable performance and is therefore adopted for subsequent experiments. To evaluate the model, we used five-fold cross-validation with disjoint image sets. In each round, four folds are designated for model development, with $80\%$ of their samples used for training and the remaining $20\%$ is used for validation. The fifth fold serves as the test set. This process is carried out across all five folds, with each fold serving once as the test set, and the final performance is reported as the mean over the five evaluation rounds.  

Our ABiG-Net is implemented with the PyTorch framework. We use the Adam optimizer for both upper- and lower-level optimization. We use a learning rate of $10^{-4}$ for the lower-level and $10^{-3}$ for the upper-level optimization, and these are obtained by fine-tuning through a grid search. For both datasets, we train our model for $200$ epochs with a batch size of $20$. 

To obtain a suitable value for L1-norm coefficient $\lambda_{\text{sparse}}$, we perform a grid search over $\{10^{-2},10^{-3},5\times10^{-4},10^{-4}\}$. We observe the best and stable performance at $10^{-4}$, thus we set $\lambda_{\text{sparse}}=10^{-4}$. The temperature parameters in Algorithm 1 are selected empirically. We set the initial temperature $\tau_{init} = 1.0$ to encourage broader exploration of adjacency structures in the early training phase. The decay rate $\gamma$ is crucial; stable performance is observed for values between $0.95$ and $0.99$. We choose $\gamma=0.98$ across all datasets. We set $\tau_{min}=0.1$ to retain moderate stochasticity in later training stages, preventing overly deterministic adjacency sampling and convergence to local minima. 

We train ABiG-Net on a single NVIDIA Tesla T4 GPU, which takes approximately $3$ hours for Dataset I, and around $4.5$ hours for Dataset II. Average inference latency per graph is $4.8$ ms (Dataset I) and $6.5$ ms (Dataset II). We use the same architecture for the image classifier and the adjacency generation network in our ABiG-Net framework for both dataset, resulting in about $0.86$M trainable parameters in total.

\subsection{Experimental Results and Analysis}

\subsubsection{Extended CRC dataset} 

To validate the effectiveness of our approach on the Extended CRC dataset (Dataset I), we benchmarked ABiG-Net against a comprehensive set of state-of-the-art (SOTA) methods, as summarized in Table~\ref{tab:extended_crc_results}. The results highlight that ABiG-Net consistently provides strong performance across both three-class and binary classification tasks.

\begin{table}[h]
\centering
\caption{Performance comparison on the Extended CRC dataset (Dataset I) under both three-class (Normal, Low-grade, High-grade) and binary (Normal vs. Cancer) classification settings. Results are reported as mean accuracy ($\pm$ standard deviation).}
\label{tab:extended_crc_results}
\renewcommand{\arraystretch}{1.4}
\setlength{\tabcolsep}{12pt} %
\begin{tabular}{|l|c|c|}
\hline
\textbf{Method} & \textbf{Three-class (\%)} & \textbf{Binary (\%)} \\
\hline
ResNet50 \cite{resnet50} & $86.33 \pm 0.94$ & $95.67 \pm 2.05$ \\ \hline
MobileNet \cite{mobilenet} & $84.33 \pm 3.30$ & $95.33 \pm 2.49$ \\ \hline
InceptionV3 \cite{inceptionv3} & $84.67 \pm 1.70$ & $92.78 \pm 2.74$ \\ \hline
Xception \cite{xception} & $86.67 \pm 0.94$ & $97.00 \pm 2.83$ \\ \hline
CA-CNN \cite{extend} & $86.67 \pm 1.70$ & $97.67 \pm 0.94$ \\ \hline
ViT \cite{vit} & $86.67 \pm 4.04$ & $96.67 \pm 1.52$ \\ \hline
CGC-Net \cite{cgcnet} & $93.00 \pm 0.93$ & $98.33 \pm 0.98$ \\ \hline
HAT-Net \cite{hatnet} & $95.33 \pm 0.58$ & $98.33 \pm 0.98$ \\ \hline
C2P-GCN \cite{c2pgcn} & $95.00 \pm 1.70$ & $98.00 \pm 1.00$ \\ \hline
\textbf{ABiG-Net (Ours)} & $\mathbf{97.33 \pm 1.15}$ & $\mathbf{98.33 \pm 0.58}$ \\
\hline
\end{tabular}
\end{table}

For the three-class classification setting, where the model is required to distinguish between normal, low-grade cancer, and high-grade cancer, our proposed ABiG-Net achieved an accuracy of $97.33 \pm 1.15\%$. This performance surpasses all CNN-based baselines, including ResNet50, MobileNet, InceptionV3, Xception, CA-CNN, and ViT. In particular, ABiG-Net demonstrates approximately $10.66\%$ absolute improvement compared to the best performing CNN-based method. When comparing to the GCN-based methods such as CGC-Net, HAT-Net, and C2P-GCN, our proposed ABiG-Net also shows clear gains: for instance, it improves upon the best GCN-based candidate (HAT-Net) by $2.0\%$. On the other hand, in the binary classification task (cancer vs. normal), ABiG-Net achieves an accuracy of $98.33 \pm 0.58 \%$, outperforming CNN-based methods and performing on par with GCN-based methods such as CGC-Net and HAT-Net. These results clearly highlight the advantage of explicitly modeling patch-level interactions through graph learning, which benefits both binary and three-class prediction tasks.
%and confirm that the benefits of graph-structured modeling hold consistently across both binary and three-class colorectal cancer (CRC) grading tasks.

\begin{figure}[h]
\centering
\includegraphics[width=7.5cm, height=4.75cm]{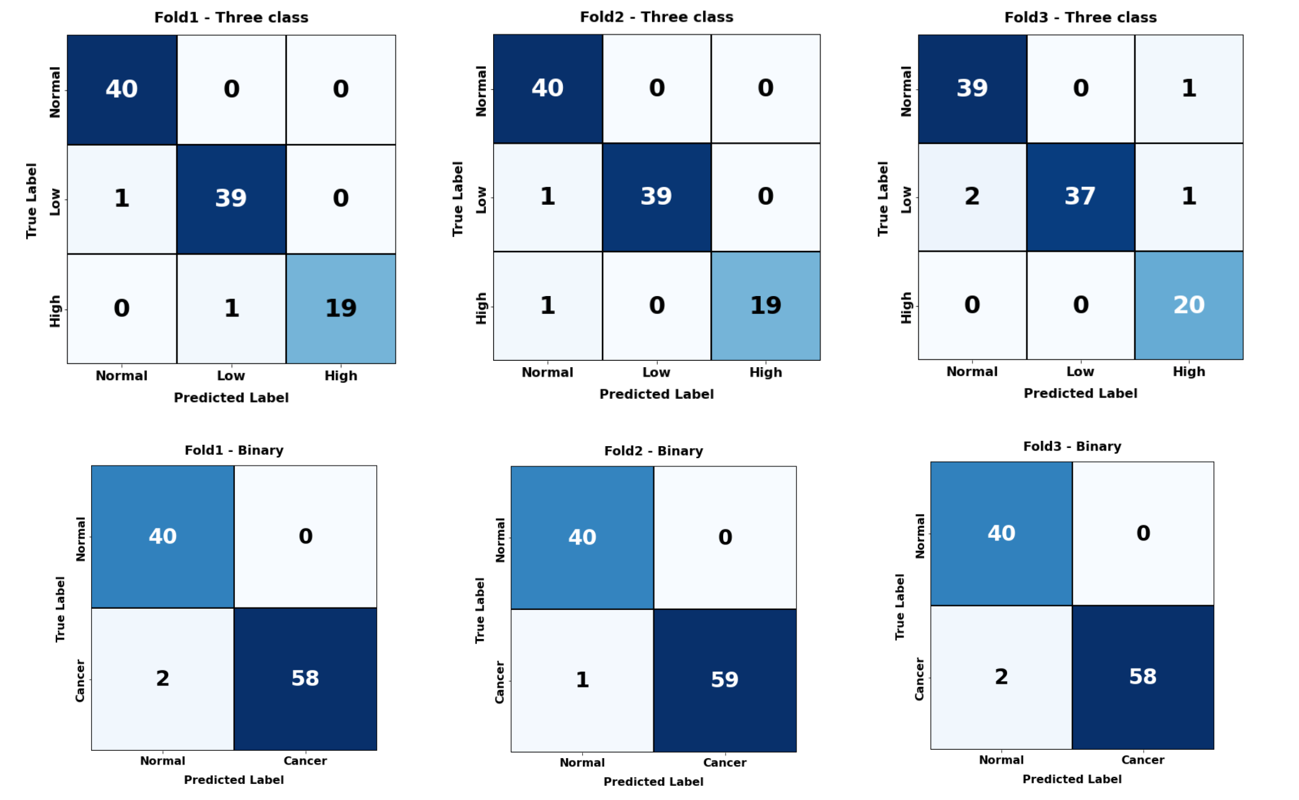}
\caption{Confusion matrices showing fold-wise classification results of ABiG-Net on the Extended CRC dataset for both three-class (top row) and binary (bottom row) settings.}
\label{fig:confusion_matrix_CRC}
\end{figure}

A closer look at the fold-wise results, illustrated in Fig.~\ref{fig:confusion_matrix_CRC} and Table~\ref{tab:fold_wise_crc}, further highlights the strengths and limitations of our model. Here, Table~\ref{tab:fold_wise_crc} shows the fold-wise accuracy and F1 scores, while Fig.~\ref{fig:confusion_matrix_CRC} presents the corresponding confusion matrices. In the three-class setting, when Fold~1 is used as the test set, the model misclassifies one low-grade cancer sample as normal and one high-grade cancer sample as low grade. For Fold~2, one low-grade cancer sample is predicted as normal, and one high-grade cancer sample is also misclassified as normal. In Fold~3, one normal tissue is predicted as high-grade, two low-grade cancer samples are misclassified as normal, and another one low-grade sample is misclassified as high-grade. 
These results indicate that the model exhibits strong reliability in identifying both normal and high-grade colorectal tissue, with only one normal and two high-grade samples misclassified across all folds. Most errors arise from low-grade cases, which are sometimes predicted as normal or, less frequently, as high-grade.

The stratification of colorectal histology images into normal, low-grade, and high-grade categories is primarily based on the degree of gland formation, with gland morphology and organization serving as key diagnostic criteria. Normal tissue preserves regular, well-formed glandular structures, whereas cancerous tissue exhibits varying levels of disruption. In the commonly used two-tiered grading system, a tumour is considered low grade if at least $50\%$ of its area retains glandular formation, and high grade if the proportion of glandular tissue falls below $50\%$ \cite{bam, bam2}. As a result, low-grade cancers often preserve much of their glandular architecture, making them resemble normal tissue, while high-grade cancers, depending on the sampled patch, may still retain partial glandular patterns or lack sufficient malignant structures, leading to confusion with low-grade cases. In the binary setting, the trends are consistent. In Fold~1, two cancer cases are misclassified as normal; in Fold~2, a single cancer case is misclassified as normal; and in Fold~3, two cancer cases are predicted as normal. Across all folds, every normal case is correctly identified with no misclassifications. These binary results reinforce the model’s strong capability to distinguish normal from cancerous tissue. The few errors are likely associated with borderline patches where malignant morphology is either subtle or underrepresented.

\begin{table}[h]
\centering
\caption{Fold-wise performance of ABiG-Net on the Extended CRC dataset under three-class and binary classification tasks, reported in terms of Accuracy and F1 score (mean $\pm$ standard deviation).}
\label{tab:fold_wise_crc}
\renewcommand{\arraystretch}{1.4}
\setlength{\tabcolsep}{4pt}
\begin{tabular}{|c|cc|cc|}
\hline
\multirow{2}{*}{\textbf{Folds}} & \multicolumn{2}{c|}{\textbf{Three-class (\%)}} & \multicolumn{2}{c|}{\textbf{Binary (\%)}} \\
\cline{2-5}
 & \textbf{Accuracy} & \textbf{F1 score} & \textbf{Accuracy} & \textbf{F1 score} \\
\hline
Fold 1 & 98.00 & 97.90 & 98.00 & 98.31 \\
Fold 2 & 98.00 & 97.91 & 99.00 & 99.16 \\
Fold 3 & 96.00 & 95.88 & 98.00 & 98.31 \\
\hline
\textbf{Overall} & \textbf{97.33 $\pm$ 1.15} & \textbf{97.23 $\pm$ 1.17} & \textbf{98.33 $\pm$ 0.58} & \textbf{98.59 $\pm$ 0.49} \\
\hline
\end{tabular}
\end{table}

\begin{figure}[h]
\centering
\includegraphics[width=8.5cm, height=3.3cm]{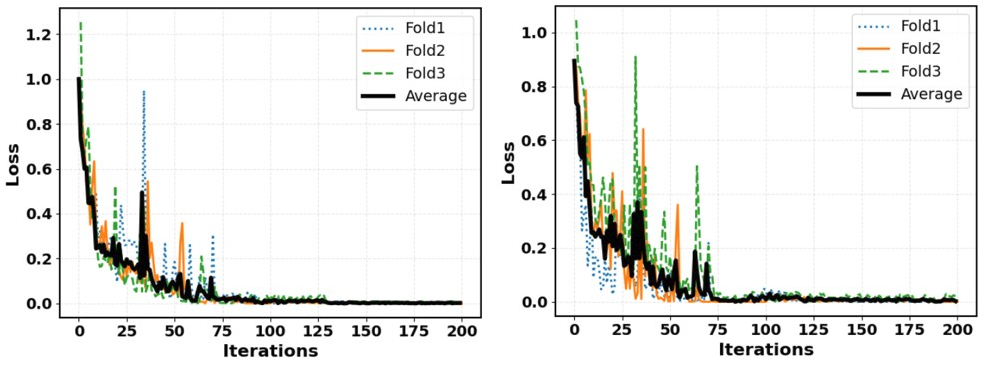}
\caption{Lower-level (left) and upper-level loss (right) across iterations for all three folds (thin colored curves) and their average (bold black) for Extended CRC dataset.}
\label{fig:convergence_crc}
\end{figure}

Fig.~\ref{fig:convergence_crc} shows the convergence plots for the lower-level loss and the upper-level loss across iterations for ABiG-Net for Dataset I. In our bilevel formulation, the lower level updates the classifier parameters $\theta$ by minimizing the training cross-entropy on $\mathcal{D}_{\text{train}}$. Concurrently, the upper level updates the adjacency-generator parameters $\psi$ by minimizing a validation objective, $\mathcal{L}_{val} + \mathcal{R}(\psi)$, which consists of the validation cross-entropy and L1-regularization. As shown in the figure, both losses decrease rapidly within the first $50$ iterations (approximately) and then stabilize near zero by around $80-100$ iterations. This indicates the successful convergence of both the classifier (left panel) and the adjacency generator (right panel). The occasional transient spikes early in training are expected due to the stochastic Gumbel-Sigmoid edge sampling and temperature annealing. As the temperature parameter $\tau$ decreases, the sampling becomes more deterministic, causing the curves to flatten. 
The fold average (bold, black line) remains smooth and closely tracks each fold, indicating stable optimization and low fold-to-fold variance.

Overall, the above results provide strong empirical evidence that our proposed ABiG-Net effectively leverages hierarchical graph structure learning to capture both local and global dependencies across tissue patches. This leads to more accurate characterization of colorectal cancer histopathology compared to traditional CNN-based pipelines and even existing GCN-based alternatives.

\subsubsection{Melanoma dataset}

We further validate the effectiveness of our approach on the Melanoma dataset (Dataset II), where the task is to classify tumor ROIs from lymphocyte-rich ROIs. ABiG-Net is benchmarked against recently proposed CNN- and GCN-based approaches, with the comparative results summarized in Table~\ref{tab:melanoma_results}.

\begin{table}[h]
\centering
\caption{Performance comparison on the Melanoma dataset (Dataset II). Results are reported as mean accuracy ($\pm$ standard deviation)}
\label{tab:melanoma_results}
\renewcommand{\arraystretch}{1.4}
\setlength{\tabcolsep}{18pt} %
\begin{tabular}{|l|c|}
\hline
\textbf{Method} & \textbf{Accuracy (\%)} \\
\hline
ResNet50 \cite{resnet50} & $94.78 \pm 2.40$ \\ \hline
MobileNet \cite{mobilenet} & $92.64 \pm 2.79$ \\ \hline
InceptionV3 \cite{inceptionv3} & $96.08 \pm 1.88$ \\ \hline
Xception \cite{xception} & $93.19 \pm 2.31$ \\ \hline
CGC-Net \cite{cgcnet} & $95.06 \pm 1.30$ \\ 
\hline
C2P-GCN \cite{c2pgcn} & $95.15 \pm 1.88$ \\ \hline
\textbf{ABiG-Net (Ours)} & $\mathbf{96.27 \pm 0.74}$ \\
\hline
\end{tabular}
\end{table}

The performance trends on this dataset are consistent with those observed on the Extended CRC dataset. ABiG-Net achieves the highest average accuracy of $96.27 \pm 0.74\%$, demonstrating both robustness and consistency. Among CNN-based baselines, InceptionV3 also delivers competitive performance ($96.08 \pm 1.88\%$) but with higher variance across folds, suggesting that our method demonstrates substantially more robust and consistent performance. Other CNN methods, such as ResNet50, MobileNet, and Xception achieve moderate accuracies, while GCN-based methods CGC-Net and C2P-GCN perform competitively. 

%These results demonstrates the advantages of our image-level graph structure learning of tissue architecture. 
%modeling of tissue structure. While patch-level CNNs rely heavily on majority voting to approximate ROI-level predictions, ABiG-Net leverages explicit modeling of patch connectivity and contextual dependencies, leading to more accurate and stable classification of tumor and lymphocyte-rich regions.

\begin{figure}[h]
\centering
\includegraphics[width=7.5cm, height=5cm]{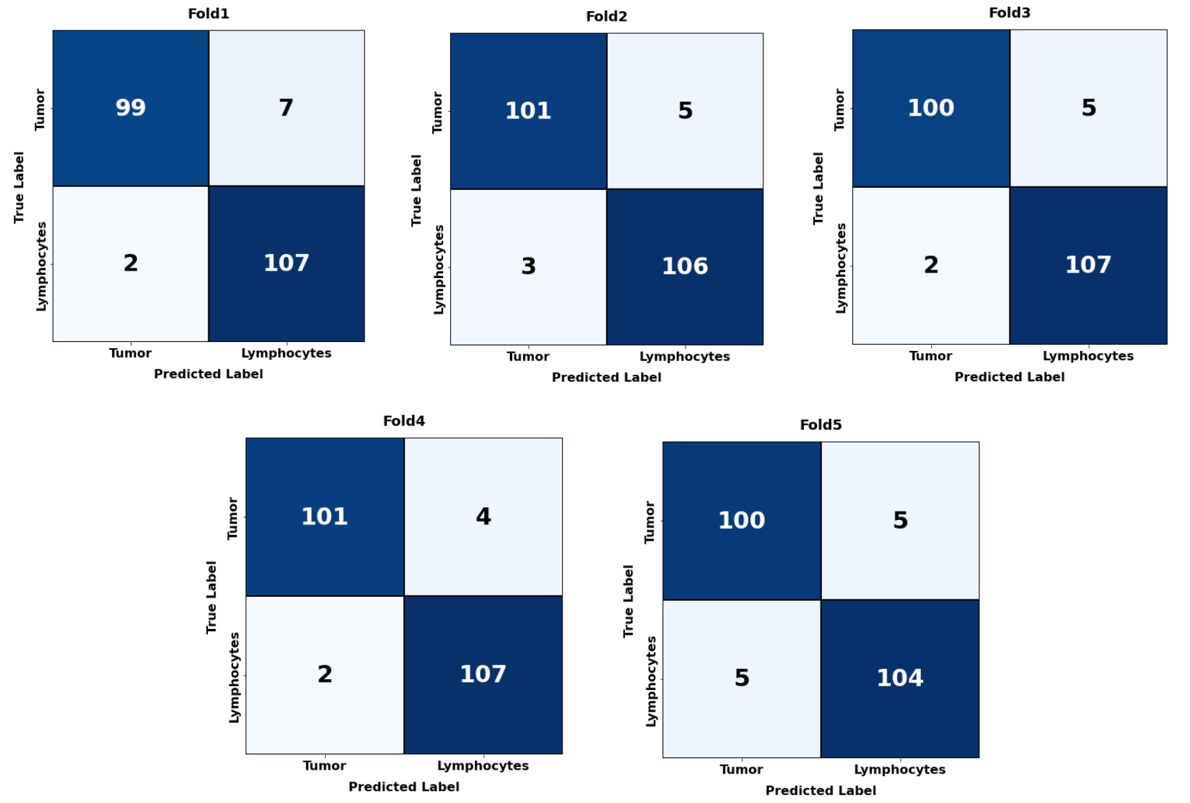}
\caption{Confusion matrices showing fold-wise classification results of ABiG-Net on the Melanoma dataset.}
\label{fig:confusion_melanoma}
\end{figure}

In Table~\ref{tab:melanoma_fold_results}, we report the fold-wise accuracy and F1 scores on the Melanoma dataset, while in Fig.~\ref{fig:confusion_melanoma} we present the corresponding confusion matrices. When considering Fold~1 as test set, the model misclassifies two lymphocyte-rich ROIs as tumor and seven tumor ROIs as lymphocytes. In Fold~2 as test set, three lymphocyte ROIs are predicted as tumor and five tumor ROIs are classified as lymphocytes. For Fold~3, two lymphocyte ROIs are misclassified as tumor and five tumor ROIs are misclassified as lymphocytes. In Fold~4, two lymphocyte ROIs are predicted as tumor and four tumor ROIs as lymphocytes. Finally, in Fold~5, five lymphocyte ROIs are misclassified as tumor and five tumor ROIs as lymphocytes. Across all five folds, the majority of both tumor and lymphocyte ROIs are classified correctly, with only a small number of misclassifications in each fold. Most errors involve tumor ROIs being misclassified as lymphocytes, although a smaller number of lymphocyte ROIs are also occasionally predicted as tumor. This reflects the intrinsic difficulty of separating tumor and lymphocyte regions in melanoma histology. Tumor ROIs are typically characterized by irregular morphology, pleomorphic nuclei, and heterogeneous spatial organization, whereas lymphocyte-rich ROIs consist of small, round, and densely clustered nuclei. In borderline regions, such as when lymphocytes infiltrate tumor tissue or tumor cells exhibit less pleomorphic morphology, these cues become less distinct, leading to occasional confusion. 

%Fig.~\ref{fig:accuracy_vs_iteration_melanoma} shows the fold-wise accuracy curves across training iterations for this dataset. The emperical results show that ABiG-Net rapidly converges within the first 100 iterations, reaching accuracies above 90%. Beyond this point, accuracy stabilizes between 95–98% across all folds, with the averaged curve (black) confirming consistent and robust performance. The small variance among folds highlights the stability and generalization of our approach.

%However, fold-wise results demonstrate that ABiG-Net achieves robust and consistent performance, with errors restricted to a limited set of challenging cases.

\begin{figure}[h]
\centering
\includegraphics[width=8.5cm, height=3.5cm]{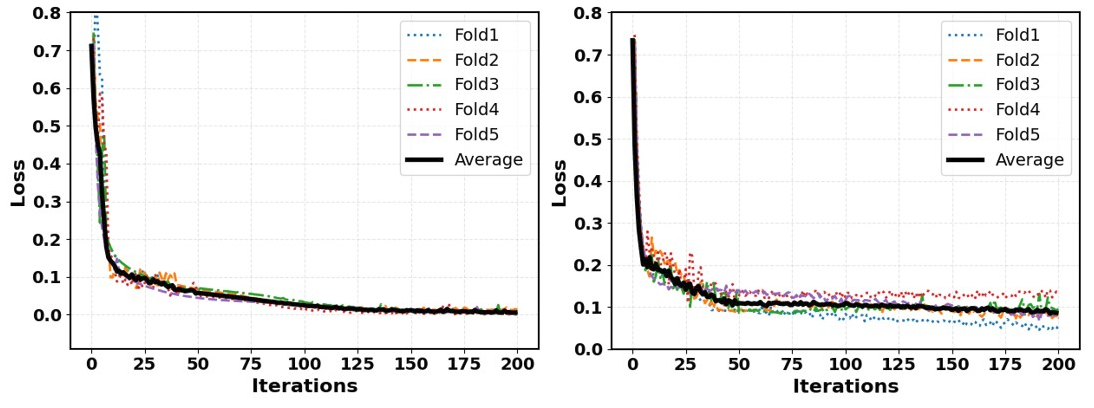}
\caption{Lower-level loss (left) and upper-level loss (right) across iterations for all five folds (thin colored curves) and their average (bold black) for Melanoma dataset.}
\label{fig:convergence_melanoma}
\end{figure}

Fig.~\ref{fig:convergence_melanoma} shows the lower-level loss and the upper-level loss across outer iterations for all five folds (thin colored curves) together with the fold-average (bold black). 
On the left, the lower-level loss drops sharply during the first $20 \text{-} 30$ iterations (approximately) and then continues a smooth decay, stabilizing near zero between the $100 \text{-} 150$ iterations across folds. This indicates that the inner objective is being optimized sufficiently over iterations. On the right, the upper-level loss decreases rapidly and then stabilizes near $0.1$ after $80 \text{-} 100$ iterations (approximately), indicating convergence of the outer‑level objective. Small fold‑specific ripples are expected due to stochastic training and edge sampling in the adjacency generator, but the fold‑average remains smooth and flat, indicating low variance across folds.

\begin{table}[h]
\centering
\caption{Fold-wise performance of ABiG-Net on the Melanoma dataset, reported in terms of accuracy and F1 score (mean $\pm$ standard deviation)}
\label{tab:melanoma_fold_results}
\renewcommand{\arraystretch}{1.3}
\setlength{\tabcolsep}{6pt}
\begin{tabular}{|c|c|c|} 
\hline
\multirow{2}{*}{\textbf{Folds}} &
  \multicolumn{2}{|c|}{\textbf{Performance (\%)}} \\ \cline{2-3}
& \textbf{Accuracy} & \textbf{F1 score} \\ \hline
Fold 1 & 95.81 & 95.65 \\ \hline
Fold 2 & 96.28 & 96.19 \\ \hline
Fold 3 & 96.73 & 96.62 \\ \hline
Fold 4 & 97.21 & 97.12 \\ \hline
Fold 5 & 95.33 & 95.24 \\ \hline
\textbf{Overall} & \textbf{96.27 $\pm$ 0.74} & \textbf{96.16 $\pm$ 0.75} \\ \hline
\end{tabular}
\end{table}

\subsection{Parameter Efficiency}

For both Dataset I and II, we use a consistent architecture for our classifier and adjacency generation networks, allowing a direct comparison of our model’s parameter efficiency against other methods as shown in Table~\ref{tab:parameters}. We note that, standard CNN-based architectures are typically benchmarked on the ImageNet dataset of $1000$ class, which requires a large classification head; our reported parameter counts are based on replacing this original head with a smaller one according to our problem. 
Although our framework is evaluated on both binary and three-class tasks, we report the adjusted counts for the three-class configuration for clarity. The parameter difference for a classification head with two and three class classification is negligible- on the order of a few thousand, which is insignificant relative to the millions of parameters in the model backbones. We therefore present the parameter counts for the three-class scenario in the table as a clear and representative measure of the parameter efficiency of our framework. As shown in the table, our ABiG-Net framework uses approximately $0.86$ M parameters, which is comparable to the highly efficient C2P-GCN ($0.85$M), but substantially lower than other prominent CNN and GCN-based methods. This significant reduction in trainable parameters underscores the parameter efficiency of our framework, which achieves competitive accuracy without relying on the massive parameterization typically required by traditional deep learning architectures.

\begin{table}[h!]
\centering
\caption{Comparison by Number of Parameters}
\label{tab:parameters}
\renewcommand{\arraystretch}{1.4}
\setlength{\tabcolsep}{14pt} %
\begin{tabular}{|l|c|}
\hline
\textbf{Method} & \textbf{\# Params (Approx.)} \\
\hline
ResNet50 & 23.6M \\
\hline
MobileNetV2 & 2.2M \\
\hline
InceptionV3 & 21.8M \\
\hline
Xception & 20.9M \\
\hline
CGC-Net & 1.44M \\
\hline
HAT-Net & 25.5M \\
\hline
C2P-GCN & 0.85M \\
\hline
\textbf{ABiG-Net} & 0.86M \\
\hline
\end{tabular}
\label{tab:param_comparison}
\end{table}

\subsection{Visual Interpretation of the Learned Graph}

\begin{figure*}[h]
  \centering
  \includegraphics[width=\textwidth]{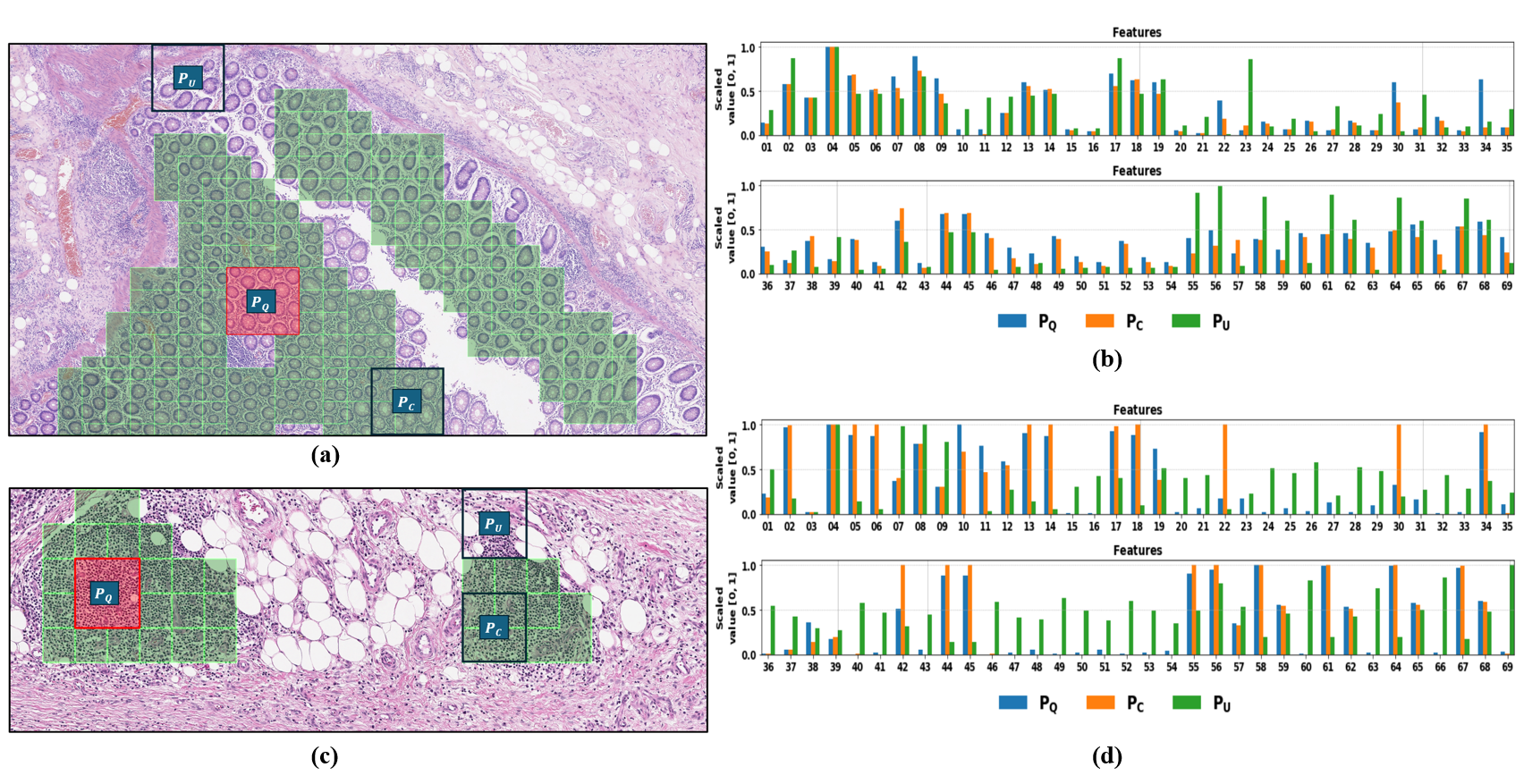}
  \caption{Visual interpretation of the learned graph for representative samples from Dataset~I (Extended CRC) and Dataset~II (Melanoma).
  (a) Learned graph for a sample image from Dataset~I: the red patch (\(P_Q\)) is the query node; green patches indicate the nodes connected to \(P_Q\) by the learned adjacency; \(P_C\) is a selected sample connected patch and \(P_U\) is a selected sample nonconnected patch.
  (b) Corresponding $69$-feature profiles for \(P_Q\), \(P_C\), and \(P_U\) (min–max normalized to \([0,1]\)); the x–axis shows Feature IDs \(1\!-\!69\) grouped by the five families (Cell-graph \(|\) Voronoi \(|\) Delaunay \(|\) MST \(|\) Nearest-neighbor) defined in Tables~\ref{tab:cell_graph_feats} and \ref{tab:patch_graph_feats}.
  (c) Learned graph for a sample image from Dataset~II and (d) its corresponding feature profiles in the same format.}
  \label{fig:visua_full}
\end{figure*}

To qualitatively assess the learned graph, we visualize the connections formed for a sample query node (red patch) in representative images from Dataset I and Dataset II, as shown in Fig.~\ref{fig:visua_full}(a) and Fig.~\ref{fig:visua_full}(c), respectively. In these examples, connected nodes are highlighted in green, providing a direct view of the relationships captured by the learned adjacency matrix. In Fig.~\ref{fig:visua_full}, $P_Q$ represents a sample query node, $P_C$ denotes a sample connected neighbor to the query node, and $P_U$ represents a randomly selected nonconnected node. 

The first example (Fig.~\ref{fig:visua_full}a) is a $5376 \times 7168$ pixel colorectal tissue image, tiled into $494$ overlapping patches of size $768 \times 768$. The query node, $P_Q$, located in a region of healthy tissue, mainly connects patches that exhibit similar, well-formed glandular structures. To examine what drives this connection, in Fig.~\ref{fig:visua_full}(b) we plot the entire $69$‑feature vector for $P_Q$ (blue bar), $P_C$ (orange bar), and $P_U$ (green bar). Features are min-max normalized to $[0,1]$ for compatibility and indexed according to Tables \ref{tab:cell_graph_feats} and \ref{tab:patch_graph_feats}. A detailed comparison of the feature profiles reveals that various features across the cell graph, Voronoi, Delaunay, MST, and nearest-neighbor families show $P_C$ closely tracking the feature profile of the query patch $P_Q$, while the unconnected patch $P_U$ diverges across most features. These features capture local cellular organization and topology, including clustering, crowding, and isolation. The connected patches demonstrate that the model correctly identifies and associates regions that share similar glandular architecture and nuclear organization, consistent with the image’s ground-truth label of normal tissue~\cite{bam}. Hence, the learned graph effectively captures the morphological homogeneity characteristic of normal colorectal tissue. 

The second example (Fig.~\ref{fig:visua_full}c) is a $1788 \times 3368$ pixel image from Dataset II, tiled into $60$ overlapping $512 \times 512$ patches. The query node, $P_Q$, encompasses a region of dense lymphoid infiltration. The learned graph connects this node to other regions, often spatially distant, that display similar lymphoid aggregation, characterized by compact, hyperchromatic nuclei and scant cytoplasm. 
The corresponding feature distributions in Fig.~\ref{fig:visua_full}(d) show the same trend observed in Dataset I. $P_C$ again aligns with $P_Q$ across multiple feature families (cell graph, voronoi, delaunay, minimum spanning tree, and nearest neighbor), whereas $P_U$ diverges markedly.  The connected patches confirm the model's ability to learn associations based on specific cytological features rather than mere spatial proximity.

Across both samples, the learned connections provide an interpretable view of how ABiG-Net comes to its decision. Instead of treating the images as isolated patches, ABiG-Net learns to associate relevant regions, even those that are spatially distant. Such behavior aligns with established diagnostic practices, where similar patterns are assessed comparatively across tissue regions. The resulting graph structure reveals which regions the model considers most important in its prediction, making the decision-making process transparent and interpretable.

\section{Ablation Study}

To better understand the contribution of individual components, we conduct an ablation study using the three-class setting of the Extended CRC dataset (Dataset I) and the binary setting of the Melanoma dataset (Dataset II). The binary CRC case is excluded because the three-class task already captures the same distinctions and provides a more comprehensive evaluation.

\subsection{Effect of Replacing the Learned Adjacency with a Fixed Graph}

We evaluated the importance of the learnable adjacency generator in ABiG-Net by replacing the learned graph with a fixed graph constructed using a deterministic rule. This variant corresponds to our previously published model, C2P-GCN~\cite{c2pgcn}, which applies an image-level adjacency structure without bilevel optimization. 
We keep everything else identical- patch extraction, patch‑level features, model architecture, training/validation/test splits, optimizer, and schedules- and only change how image‑level edges are formed. Here, in the image-level graph, we connect patches whenever the cosine similarity between patch features exceeds a predefined threshold of $0.8$ for both Dataset I and II. 

This change lowers accuracy in both datasets, as shown in Table~\ref{tab:ablation_learn_graph}. On the Extended CRC dataset (three‑class), accuracy drops from $97.33 \pm 1.15\%$ with ABiG‑Net to $95.00 \pm 1.70\%$ with the fixed graph. On the Melanoma dataset, accuracy decreases from $96.27 \pm 0.74\%$ to $95.15 \pm 1.88\%$. The fixed graph also shows a larger variance across splits, suggesting reduced stability. 
In ABiG-Net, since the adjacency is updated under the outer (validation) objective in our bilevel setup, edge patterns are directly shaped to improve generalization, rather than just training for a fit. In contrast, a single global similarity threshold used in a fixed image-level graph cannot adapt to WSI or ROI-specific morphology or to shifts in feature scales, and it tends to either under-connect or over-connect patches.

\begin{table}[!h]
\centering
\caption{Classification accuracy (\%) with a learned graph (ABiG-Net) versus a fixed graph (C2P-GCN). Results are mean $\pm$ standard deviation.}
\label{tab:ablation_learn_graph}
\renewcommand{\arraystretch}{1.4}
\begin{tabular}{|l|c|c|}
\hline
\textbf{Approach} & \textbf{Dataset I (3-class)} & \textbf{Dataset II} \\
\hline
ABiG-Net & $97.33 \pm 1.15$ & $96.27 \pm 0.74$ \\
\hline
Fixed image-level     & $95.00 \pm 1.70$ & $95.15 \pm 1.88$ \\
graph (C2P-GCN) & & \\
\hline
\end{tabular}
\end{table}

\subsection{Variation in Patch-size}

The choice of patch size is a critical hyperparameter in histopathology analysis, as it defines the trade-off between architectural context and cytological detail. The patch should be large enough to capture sufficient architectural context, but small enough to preserve fine-grained cellular features without having excessive unnecessary information from irrelevant tissue regions. To assess its impact, we evaluated ABiG-Net using three different patch sizes: $512 \times 512$, $768 \times 768$, and $1024 \times 1024$ pixels, while keeping the stride fixed at 256 pixels. The results presented in Table~\ref{tab:ablation_patch} reveal that the optimal patch size and its impact are highly dependent on the nature of the dataset and the underlying histology.

For Dataset I (colorectal cancer), the model's performance is notably sensitive to patch size. It achieves its best performance of $97.33 \pm 1.15\%$ with $768 \times 768$ patches, with the larger $1024 \times 1024$ patches performing slightly lower but comparably. However, there is a significant performance degradation with $512 \times 512$ patches. This reduction can be attributed to the smaller patch size, which loses glandular architectural patterns that are essential for distinguishing normal, low-grade, and high-grade colorectal cancer regions.

On the other hand, for Dataset II, the model achieves its best performance with the smallest patch size of $512 \times 512$. Compared to the significant performance degradation observed on Dataset I between patch sizes, the accuracy on Dataset II remains broadly consistent, with a variation of only about one percentage point. This robustness reflects the distinct and localized nature of tumor and lymphocyte-rich regions. Lymphocyte-rich ROIs are characterized by dense clusters of small, hyperchromatic nuclei that are adequately captured even in $512 \times 512$ patches, while tumor ROIs display nuclear atypia and irregular organization, providing strong diagnostic cues at any scale. Larger patches include more tissue but add little additional discriminative information, resulting in comparable performance between patch sizes. These findings emphasize that the optimal patch size is not universal but must be tailored to the histological characteristics of each dataset.

\begin{table}[!h]
\centering
\caption{Impact of varying patch size on ABiG-Net classification accuracy (\%). Results are shown as mean $\pm$ standard deviation.}
\label{tab:ablation_patch}
\renewcommand{\arraystretch}{1.4}
\begin{tabular}{|l|c|c|}
\hline
\textbf{Patch size} & \textbf{Dataset I (3-class)} & \textbf{Dataset II} \\
\hline
$512 \times 512$ & $92.67 \pm 2.54$ & $96.27 \pm 0.74$ \\
\hline
$768 \times 768$    & $97.33 \pm 1.15$ & $95.52 \pm 1.88$ \\
\hline
$1024 \times 1024$    & $96.67 \pm 1.70$ & $95.25 \pm 1.26$ \\
\hline
\end{tabular}
\end{table}

\subsection{Effect of Cell Graph Neighborhood Radius}

In the patch-level cell graph, the threshold $d_p$ 
specifies the neighborhood radius that controls the Euclidean distance between nuclei centroids for defining an edge. Smaller $d_p$ values restrict connections to only immediate neighbors, resulting in sparse graphs that may underrepresent true cellular neighborhoods. Larger values, on the other hand, connect cells across broader distances, producing dense graphs that risk merging unrelated regions and obscuring local structural organization. To evaluate the effect of this parameter, we varied $d_p \in \{32, 64, 128, 256\}$ pixels while keeping all other settings fixed. The classification results are reported in Table~\ref{tab:ablation_dp}.

\begin{table}[!h]
\centering
\caption{Impact of varying $d_p$ on ABiG-Net classification accuracy (\%). Results are shown as mean $\pm$ standard deviation.}
\label{tab:ablation_dp}
\renewcommand{\arraystretch}{1.4}
\begin{tabular}{|l|c|c|}
\hline
\textbf{dp} & \textbf{Dataset I (3-class)} & \textbf{Dataset II} \\
\hline
$ 32 $ & $92.67 \pm 1.53$ & $95.98 \pm 1.07$ \\
\hline
$64$    & $97.33 \pm 1.15$ & $96.27 \pm 0.74$ \\
\hline
$128$    & $96.33 \pm 1.70$ & $95.71 \pm 1.01$ \\
\hline
$256$    & $90.67 \pm 1.53$ & $95.24 \pm 1.24$ \\
\hline
\end{tabular}
\end{table}

For Dataset I, performance is strongly dependent on the choice of $d_p$. The highest accuracy of $97.33 \pm 1.15\%$ is achieved at $d_p = 64$, suggesting this neighborhood size best preserves glandular architecture while minimizing noise. This observation aligns with prior studies \cite{hatnet, c2pgcn, cgcnet}, which also employed $d_p=64$ on the same dataset. When $d_p$ is too small (32), the patch-level graphs become overly sparse, breaking apart glandular structures into incomplete neighborhoods. 
In normal tissue, this loss of glandular continuity causes patches to resemble those from low- or high-grade cancers, thereby reducing accuracy significantly ($92.67 \pm 1.53\%$). A moderate increase to $d_p=128$ still yields high performance $(96.33 \pm 1.70 \%)$, suggesting that the model remains robust within a certain range of neighborhood radii. However, when $d_p$ is too large (256), the graphs become excessively dense, introducing spurious long-range cellular connections within the patch that dilute discriminative local structure, leading to degraded performance ($90.67 \pm 1.53\%$). These results highlight that for histologies in which the architectural organization of the glands is the dominant diagnostic signal, careful tuning of $d_p$ is critical.

In contrast, Dataset II shows comparable performance across all tested thresholds, with accuracies ranging narrowly between $95.24\%$ and $96.27\%$. This robustness reflects the dataset’s reliance on cytological features, such as dense lymphocyte aggregates and tumor cell atypia. These signals are inherently local and remain reliably captured regardless of the exact graph density. Even at $d_p = 32$, the graphs retain enough cellular interactions to represent these strong cytological patterns, and increasing the neighborhood radius does not provide major discriminative benefit.

Overall, these findings indicate that the optimal choice of $d_p$ depends on the dominant histological characteristics of the dataset: tasks that depend on glandular architecture (Dataset I) are highly sensitive to the neighborhood radius, whereas tasks driven by cytological signals (Dataset II) remain robust across a wide range of values.

\section{Conclusion and Future Work}

We introduced ABiG‑Net, a hierarchical graph learning framework that couples local cellular organization with global, image‑level context for histopathology image classification. Our approach is distinguished by its hierarchical representation of tissue. At the patch level, nuclei are detected and used to build complementary cell‑centric graphs (cell graph, Voronoi, Delaunay, MST, and nearest‑neighbor descriptors) from which descriptive features are derived. At the image level, a parametric adjacency generator produces sparse, symmetric connectivity among patches via a Gumbel–Sigmoid reparameterization, and a multilayer GCN computes patch embeddings for image‑level prediction. The two modules are optimized in a first‑order, approximate bilevel scheme where the classifier parameters are learned by minimizing the training loss (lower level), and adjacency parameters are learned by minimizing the validation loss with an L1 sparsity penalty (upper level). This decoupled optimization learns long-range interactions that are task-dependent and produces explicit and interpretable graphs. Our comprehensive experiments on two distinct datasets, colorectal cancer grading (Dataset I) and melanoma tumor/lymphocyte classification (Dataset II), demonstrate the superiority of our approach. ABiG-Net consistently outperforms both traditional CNN-based methods and state-of-the-art GCN models that rely on fixed graphs. The visual interpretability of the learned graphs, which correctly associate spatially distant but histologically similar regions, further validates our model's ability to emulate diagnostic reasoning. Moreover, our ablation studies empirically confirm the importance of a learned graph structure and highlight how the optimal hyperparameters for patch-level analysis are intrinsically linked to the scale of the diagnostic features in the underlying histology.

Building on the success of this framework, we envision several key directions for future work. First, we will advance the framework by learning the patch-level cell graphs directly from the data, replacing the current hand-crafted feature extraction pipeline. This would create a fully end-to-end, hierarchical learning system where both local and global tissue representations are optimized. Second, we plan to implement and evaluate a second-order approximation for the bilevel optimization. While our first-order approach proved highly effective, a more precise second-order method could potentially refine the graph learning process, and quantifying the trade-off between its computational cost and any performance gains is a critical next step. Finally, we aim to extend the application of ABiG-Net to a broader range of clinical challenges. This includes validating its performance on histopathology from other organs and, more importantly, adapting the framework to more clinically significant tasks beyond classification, such as patient survival analysis and treatment response prediction.

In summary, ABiG-Net demonstrates that learning the graph structure directly, rather than relying on fixed heuristics, leads to both higher accuracy and better interpretability in histopathology image classification. By integrating cell-level information with image-level context, the framework captures biologically meaningful relationships across tissue regions and produces graphs that are transparent to inspect. These results suggest that adaptive graph learning can provide a practical and reliable foundation for digital pathology applications, bridging the gap between computational models and diagnostic reasoning.

\section*{Acknowledgment}

The authors are grateful to the Center for Computational Innovations (CCI) at Rensselaer Polytechnic Institute for the computational support. 

\section*{CONFLICTS OF INTEREST}
The authors declare no conflicts of interest related to this work.

%The preferred spelling of the word ``acknowledgment'' in American English is
%without an ``e'' after the ``g.'' Use the singular heading even if you have
%many acknowledgments. Avoid expressions such as ``One of us (S.B.A.) would
%like to thank $\ldots$ .'' Instead, write ``F. A. Author thanks $\ldots$ .'' In most
%cases, sponsor and financial support acknowledgments are placed in the
%unnumbered footnote on the first page, not here.

% ADD THESE TWO LINES INSTEAD
%\bibliographystyle{IEEEtran}
%\bibliography{bibliography}
\bibliographystyle{IEEEtran}
\bibliography{bibliography}

\begin{IEEEbiography}[{\includegraphics[width=1in,height=1.25in,clip,keepaspectratio]{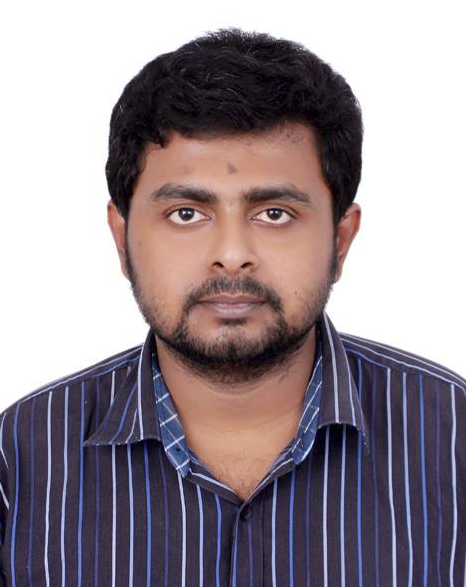}}]{Sudipta Paul} (Member, IEEE) received the B.Sc. degree in electrical and electronic engineering from Khulna University of Engineering and Technology (KUET), Khulna, Bangladesh, and the M.Sc. degree in electrical engineering from the University of Minnesota, Duluth, MN, USA. He is currently pursuing the Ph.D. degree in electrical engineering with the Department of Electrical, Computer, and Systems Engineering at Rensselaer Polytechnic Institute (RPI), Troy, NY, USA.

His research interests include machine learning, data science, control theory, medical imaging, and computational methods for disease diagnosis. He is a member of the Data Science Research Lab at RPI, where he is supervised by Prof. B\"ulent Yener.
\end{IEEEbiography}

\begin{IEEEbiography}[{\includegraphics[width=1in,height=1.25in,clip,keepaspectratio]{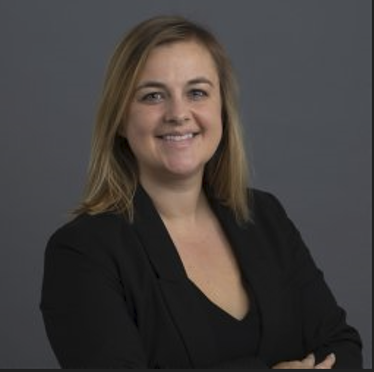}}]{AMANDA W. LUND} received the Ph.D. degree in 2007 from the Rensselaer Polytechnic Institute, Troy, NY, USA. She completed postdoctoral training at the Ecole Polytechnique Federale de Lausanne, Laboratory of Lymphatic and Cancer Bioengineering. She is currently an Associate Professor in the Ronald O. Perelman Department of Dermatology and the Department of Pathology at the NYU Grossman School of Medicine, New York, NY, USA.

Her research focuses on understanding the functional consequences of lymphatic transport and remodeling on melanoma metastasis and immune surveillance. Her laboratory applies high-content imaging, transcriptomics, and mouse models to define how lymphatic function is shaped by the tumor microenvironment and how this contributes to tumor progression. The long-term goal of her work is to identify novel targets and strategies that leverage the lymphatic vasculature to improve cancer immunotherapy.

\end{IEEEbiography}

\begin{IEEEbiography}[{\includegraphics[width=1in,height=1.25in,clip,keepaspectratio]{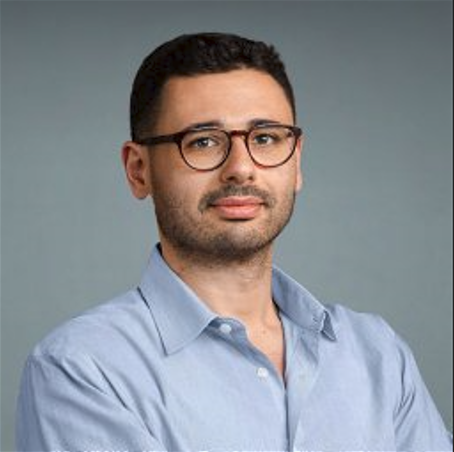}}]{George Jour} received his M.D. degree and completed residency training in anatomic pathology at St. Luke’s–Roosevelt Hospital Center, New York, NY, USA, in 2013. He subsequently completed fellowships in bone pathology at the University of Washington School of Medicine in 2014, molecular genetics at Memorial Sloan Kettering Cancer Center in 2015, and dermatopathology at the University of Texas Medical School at Houston in 2016. He is currently a Clinical Professor in the Department of Pathology and the Ronald O. Perelman Department of Dermatology at the NYU Grossman School of Medicine, New York, NY, USA. He also serves as Director of the Molecular Pathology Fellowship Program and Associate Director of the Molecular Pathology Laboratory.

His research interests include molecular and computational pathology, melanoma biology, and the development of AI-driven approaches for predicting response to immunotherapy and targeted therapies. He has co-authored numerous peer-reviewed publications in leading journals such as Clinical Cancer Research, Modern Pathology, and Cancer Research.
\end{IEEEbiography}

\begin{IEEEbiography}[{\includegraphics[width=1in,height=1.25in,clip,keepaspectratio]{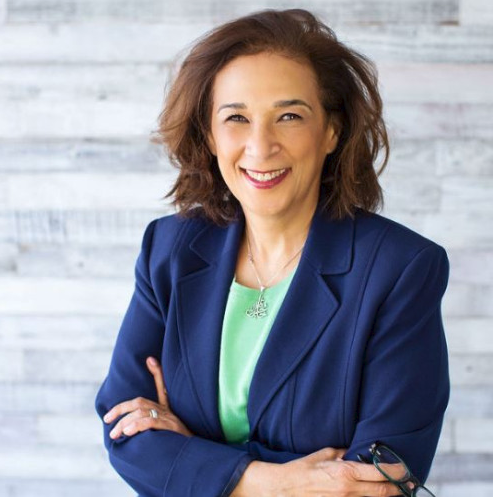}}]{Iman Osman} received the M.B.B.Ch. degree in medicine and surgery in 1982 and the M.S. degree in internal medicine in 1986 from Cairo Medical School, Cairo, Egypt, and the M.D. degree in medical oncology in 1991 from the National Cancer Institute, Cairo, Egypt. She is currently the Rudolf L. Baer Professor of Dermatology with tenure at the NYU Grossman School of Medicine, New York, NY, USA, where she also holds professorships in the Departments of Medicine and Urology. She serves as Associate Dean for Clinical Research Strategy and directs the Interdisciplinary Melanoma Cooperative Group (IMCG), a large translational research program she founded in 2002.

Her research focuses on elucidating the biological processes that drive melanoma’s aggressive clinical behavior, identifying novel molecular targets for therapy, and clarifying the links between treatment response and toxicity. Her team develops simple, clinically applicable tools and assays using advanced technologies, and applies machine learning to generate predictive models from multi-dimensional patient data.
\end{IEEEbiography}

\begin{IEEEbiography}[{\includegraphics[width=1in,height=1.25in,clip,keepaspectratio]{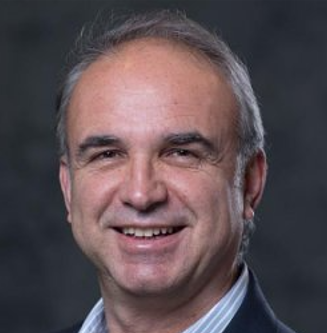}}]{B\"ulent Yener} (Fellow, IEEE) received the M.S. and Ph.D. degrees in computer science from Columbia University, New York, NY, USA, in 1987 and 1994, respectively. He is currently a Professor in the Department of Computer Science and in the Department of Electrical, Computer, and Systems Engineering at Rensselaer Polytechnic Institute (RPI), Troy, NY, USA, where he also serves as the Founding Director of the Data Science Research Center. Before joining RPI, he was a Member of the Technical Staff at Bell Laboratories, Murray Hill, NJ, USA.

Dr. Yener’s research spans computer communications networks, information security and privacy, biomedical informatics, and artificial intelligence. He is well known for pioneering the \textbf{cell-graph} methodology for modeling structure–function relationships in tissue, which has been widely adopted in digital pathology. His work integrates combinatorial optimization, machine learning, and data science to address problems across engineering and biomedicine.

He was a Marie Curie Fellow from 2009 to 2010 and was named a Fellow of the IEEE in 2015 for contributions to network design optimization and security.
\end{IEEEbiography}

\newpage

%\EOD

\end{document}